\documentclass[aps,prr,twocolumn,amsmath,amssymb,floatfix,superscriptaddress,citeautoscript]{revtex4-2}

\usepackage{graphicx}
\usepackage{natbib}
\usepackage{hyperref}

\usepackage{comment}
\usepackage[percent]{overpic}

\newcommand{\bea}{\begin{eqnarray}}
\newcommand{\I}{\text{i}}
\newcommand{\eea}{\end{eqnarray}}

\newcommand{\bx}{\mathbf{x}}

\newcommand{\bk}{\mathbf{k}}

\newcommand{\bp}{\mathbf{p}}

\newcommand{\bM}{\mathbf{M}}

\newcommand{\bq}{\mathbf{q}}

\newcommand{\cH}{\mathcal{H}}

\newcommand{\tq}{\tilde{q}}

\begin{document}

\title{Topological features in the ferromagnetic Weyl semimetal CeAlSi: \\ Role of domain walls}

\author{M. M. Piva}
\email{Mario.Piva@cpfs.mpg.de}
\affiliation{Max Planck Institute for Chemical Physics of Solids, N\"{o}thnitzer Str.\ 40, D-01187 Dresden, Germany}

\author{J. C. Souza}
\altaffiliation[Present address: ]
{Department of Condensed Matter Physics, Weizmann Institute of Science, Rehovot, Israel.}
\affiliation{Instituto de F\'{\i}sica ``Gleb Wataghin'', UNICAMP, 13083-859, Campinas, SP, Brazil}
\affiliation{Max Planck Institute for Chemical Physics of Solids, N\"{o}thnitzer Str.\ 40, D-01187 Dresden, Germany}

\author{V. Brousseau-Couture}
\affiliation{D\'{e}partement de Physique, Universit\'{e} de Montr\'{e}al, C.P. 6128, Succursale Centre-Ville, Montr\'{e}al, Qu\'{e}bec, Canada H3C 3J7}

\author{Sopheak Sorn}
\affiliation{Institute for Quantum Materials and Technology,
Karlsruhe Institute of Technology, 76131 Karlsruhe, Germany}

\author{K. R. Pakuszewski}
\affiliation{Instituto de F\'{\i}sica ``Gleb Wataghin'', UNICAMP, 13083-859, Campinas, SP, Brazil}

\author{Janas K. John}
\affiliation{Max Planck Institute for Chemical Physics of Solids, N\"{o}thnitzer Str.\ 40, D-01187 Dresden, Germany}

\author{C. Adriano}
\affiliation{Instituto de F\'{\i}sica ``Gleb Wataghin'', UNICAMP, 13083-859, Campinas, SP, Brazil}

\author{M. C\^{o}t\'{e}}
\affiliation{D\'{e}partement de Physique, Universit\'{e} de Montr\'{e}al, C.P. 6128, Succursale Centre-Ville, Montr\'{e}al, Qu\'{e}bec, Canada H3C 3J7}

\author{P. G. Pagliuso}
\affiliation{Instituto de F\'{\i}sica ``Gleb Wataghin'', UNICAMP, 13083-859, Campinas, SP, Brazil}
\affiliation{Los Alamos National Laboratory, Los Alamos, New Mexico 87545, USA}

\author{Arun Paramekanti}
\affiliation{Department of Physics, University of Toronto, 60 St. George Street, Toronto, ON, M5S 1A7 Canada}
\affiliation{S. N. Bose National Centre for Basic Sciences, Block JD, Sector - III, Salt lake, Kolkata-700106, India}

\author{M. Nicklas}
\email{Michael.Nicklas@cpfs.mpg.de}
\affiliation{Max Planck Institute for Chemical Physics of Solids, N\"{o}thnitzer Str.\ 40, D-01187 Dresden, Germany}

\date{\today}

\begin{abstract}

In the ferromagnetic (FM) Weyl semimetal CeAlSi both space-inversion and time-reversal symmetries are broken. Our quantum oscillation (QO) data indicate that the FM ordering modifies the Fermi surface topology and also leads to an unusual drop in the QO amplitude. In the FM phase, we find a pressure-induced suppression of the anomalous and the loop Hall effects. This cannot be explained based on the electronic band structure or magnetic structure, both of which are nearly pressure independent. Instead, we show that a simplified model describing the scattering of Weyl fermions off FM domain walls can potentially explain the observed topological features. Our study highlights the importance of domain walls for understanding transport in FM Weyl semimetals.

\end{abstract}

\maketitle

\section{INTRODUCTION}

Topological phases of matter have lately received considerable attention, due to the experimental realization of exotic types of charge carriers. One example is the massless  Weyl fermions found in Weyl semimetals (WSMs) \cite{yan2017topological,zhang2018towards,armitage2018weyl}, which are characterized by remarkable electronic properties, such as surface Fermi arcs,  a bulk chiral anomaly, axial–gravitational anomaly, an extremely large magnetoresistance (MR) and an anomalous Hall effect (AHE) \cite{yan2017topological,armitage2018weyl,yang2018symmetry,GoothNature2017,wawrzikPRL2021}. Weyl fermions can be generated by either breaking space-inversion (SI) or time-reversal (TR) symmetry of materials with a Dirac or quadratic band touching points. So far most experimentally studied WSMs break SI symmetry \cite{weng2015weyl,huang2015weyl,xu2015discovery,zhang2016signatures,huang2015observation,shekhar2015extremely,ali2014large,zhu2015quantum,jiang2017signature}; fewer examples are known for WSMs with broken TR symmetry, i.e.\ \textit{magnetic} WSMs \cite{kuroda2017evidence,yang2017topological,morali2019fermi,liu2019magnetic,belopolski2019discovery,borisenko2019time}. Magnetic WSMs are of fundamental interest since they intertwine topology and strong correlations \cite{grefe2020weyl,paschen2021quantum}. They also offer the potential to manipulate the topological phase in a desired way, for instance using a magnetic field to tune the position of Weyl nodes or to control the chirality or geometry of magnetic domain walls, which is important for next-generation spintronics applications \cite{kurebayashi2016voltage,yang2021chiral}.

The family of $Ln$Al$Pn$ ($Ln$ = lanthanides, $Pn$ = Ge, Si) materials is ideal to host nontrivial topological properties due to their noncentrosymmetric crystalline structure ($I4_{1}md$), which is the same as in the TaAs family of WSMs \cite{weng2015weyl,xu2015experimental,yang2015weyl,xu2015discovery,arnold2016negative,liu2016evolution}. Multiple Weyl nodes and a large spin Hall effect were predicted to exist in LaAlGe and LaAlSi \cite{ng2021origin}. Weyl cones were experimentally observed for LaAlGe \cite{xu2017discovery} and a $\pi$~Berry phase was recently found in LaAlSi \cite{su2021multiple}. Remarkably, magnetic members of the family host rare-earth moments which can order and additionally break TR symmetry - many of them are predicted to feature Weyl nodes near the Fermi level \cite{yang2020transition,xu2021shubnikov,chang2018magnetic,yang2021noncollinear}.
Experiments have discovered an anomalous Hall effect (AHE) in PrAlGe$_{1-x}$Si$_{x}$ \cite{yang2020transition}, chiral surface Fermi arcs in PrAlGe \cite{sanchez2020observation,destraz2020magnetism}, and a topological magnetic phase and singular angular MR in the semimetal CeAlGe \cite{hodovanets2018single,suzuki2019singular,puphal2020topological}. In addition, Weyl fermions have been found to mediate magnetism in NdAlSi \cite{gaudet2020incommensurate} and a $\pi$~Berry phase was reported for quantum oscillations (QO) in SmAlSi \cite{xu2021shubnikov}.

In this Article, we focus on the ferromagnetic Weyl semimetal CeAlSi. CeAlSi, which hosts an in-plane non-collinear ferromagnetic (FM) order below the Curie temperature $T_C \approx 8$~K with a large anisotropy, the $c$-axis being the magnetically hard axis \cite{yang2021noncollinear}. Ce$^{3+}$ spins in adjacent FM planes display an angle of $\approx 70^\circ$ \cite{yang2021noncollinear}. Recent angle resolved photo emission spectroscopy experiments in the paramagnetic phase of CeAlSi above $T_C$ revealed Fermi arcs and several Weyl nodes lying close to the Fermi energy which stem from the non-centrosymmetric structure \cite{sakhya2022observation}. Going below $T_C$, into the FM state, a magnetic field applied parallel to the $[100]$ direction reveals an AHE, while a $[001]$ field leads to an unexplained hysteretic loop Hall effect (LHE) \cite{yang2021noncollinear}. In addition, CeAlSi may exhibit nontrivial magnetic domain walls \cite{huang2021topological}; indeed, chiral domain walls were recently detected in this system \cite{sun2021mapping}. Furthermore,  magnetoelastic couplings give rise to picometer displacements in the unit cell due to the internal FM field, which can lead to different domain wall spin textures \cite{xu2021picoscale}. 

The presence of this magnetoelastic effect suggests that external pressure may lead to a strong tuning of magnetism and to associated large changes in the AHE and LHE \cite{xu2021picoscale}. Hydrostatic pressure has previously been shown to be an effective tool in tuning the electronic structure without introducing any additional disorder and was successfully used to tune Weyl points closer to the Fermi energy in certain topological materials \cite{dos2016pressure,liang2017pressure,hirayama2015weyl,rodriguez2020two}. Furthermore, application of pressure is known to systematically  modify the magnetic properties in Ce-based materials \cite{nicklas2015pressure}.
Here, we use hydrostatic pressure as a tool to investigate the origin of the features characteristic of the nontrivial topological behavior in CeAlSi, focusing on longitudinal and Hall transport experiments and on quantum oscillation measurements. We combine these with {\it ab initio} density functional theory (DFT) calculations and phenomenological
models for scattering of Weyl fermions off magnetic domain walls to shed light on our unusual observations.

\section{METHODS}
\label{Methods}

Single crystals of CeAlSi and LaAlSi were grown by the Al-flux technique similar to \cite{bobev2005ternary}. High purity elements with starting composition Ce~[La]~(99.99\%) : Al~(99.999+\%) : Si~(99.999+\%), $1:20:1$, were place into an alumina crucible and sealed in an evacuated quartz tube. The samples were heated to $1200^{\circ}$C, kept at this temperature for 15 hours and cooled down to $720^{\circ}$C at $2^{\circ}$C/h. The excess of Al was removed by spinning the tube upside down in a centrifuge. The crystal structure was confirmed by x-ray powder diffraction. Energy dispersive x-ray spectroscopy shows, within the experimental uncertainty, a Ce:Al:Si proportion of $1:1:1$.

Electrical transport experiments were carried out by a four-probe configuration using a low-frequency AC resistance bridge. Temperatures down to 1.8~K and magnetic fields up to 9~T were achieved in a physical property measurement system (PPMS, Quantum Design) and in a liquid helium cryostat (Janis). Magnetization measurements were conducted in a magnetic property measurement system (MPMS, Quantum Design). Pressures up to 2.7~GPa (electrical transport) and 1~GPa (magnetization) were generated using self-contained piston-cylinder-type pressure cells using silicon oil as pressure transmitting medium. A piece of lead (tin) served as manometer.

Density functional theory (DFT) calculations were performed with the local density approximation functional (LDA) and projector-augmented wave (PAW) method as implemented in the \textsc{Abinit} software package~\cite{gonze_abinitproject:_2019}, using Jollet-Torrent-Holzwarth (JTH) pseudopotentials~\cite{jollet_generation_2014}.  Spin-orbit coupling (SOC) and non-collinear magnetism are taken into account. An on-site Coulomb interaction with $U=6$~eV was added for the Ce $f$ electrons within the LDA+U scheme. We use a \mbox{$16\times16\times16$} Monkhorst-Pack \mbox{$\mathbf{k}$-point} grid and a plane-wave energy cutoff of 25~hartree. The lattice parameters and relevant internal atomic coordinates were optimized at respectively $0$~GPa and $3$~GPa until all forces on the atoms were below \mbox{$10^{-6}$ hartree/bohr$^3$}. At 0~GPa (3~GPa), we obtain \mbox{$a=7.926$}~bohr (\mbox{$a=7.832$}~bohr) and \mbox{$c=27.397$}~bohr (\mbox{$c=27.192$}~bohr). 

\section{RESULTS}

\subsection*{Temperature -- pressure phase diagram}

\begin{figure}[!b]
	\includegraphics[width=0.98\linewidth]{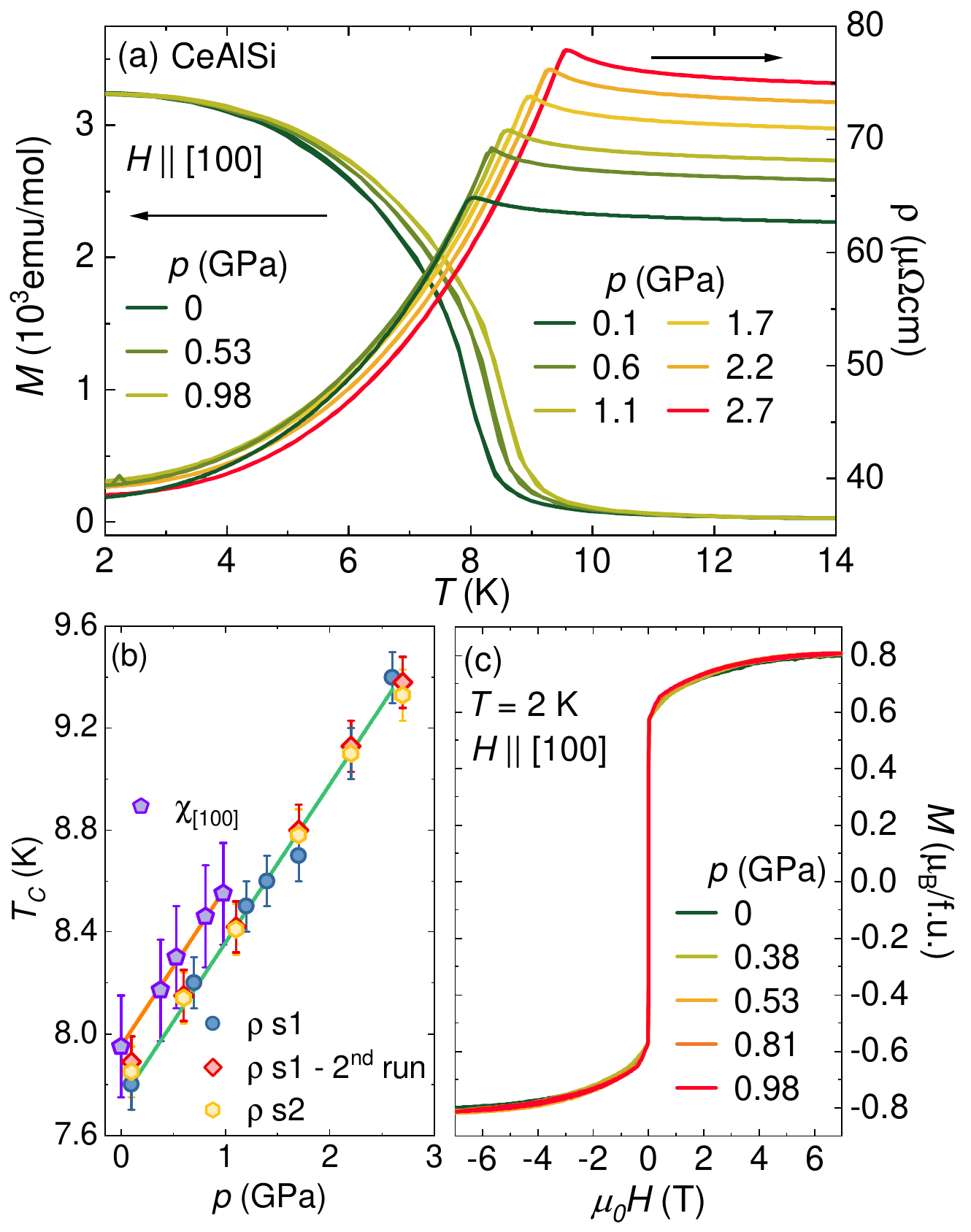}
	\caption{(a) Magnetization  ($M$) (left axis), obtained in an applied field of 50~mT along the $[100]$ crystal axis, and electrical resistivity (right axis) as a function of temperature for selected pressures. (b) Temperature--pressure phase diagram. The solid lines are linear fits. (c) Magnetization measurements for several pressures.}
	\label{chi_rho}
\end{figure}

\begin{figure*}[!t]
	\includegraphics[width=0.7\linewidth]{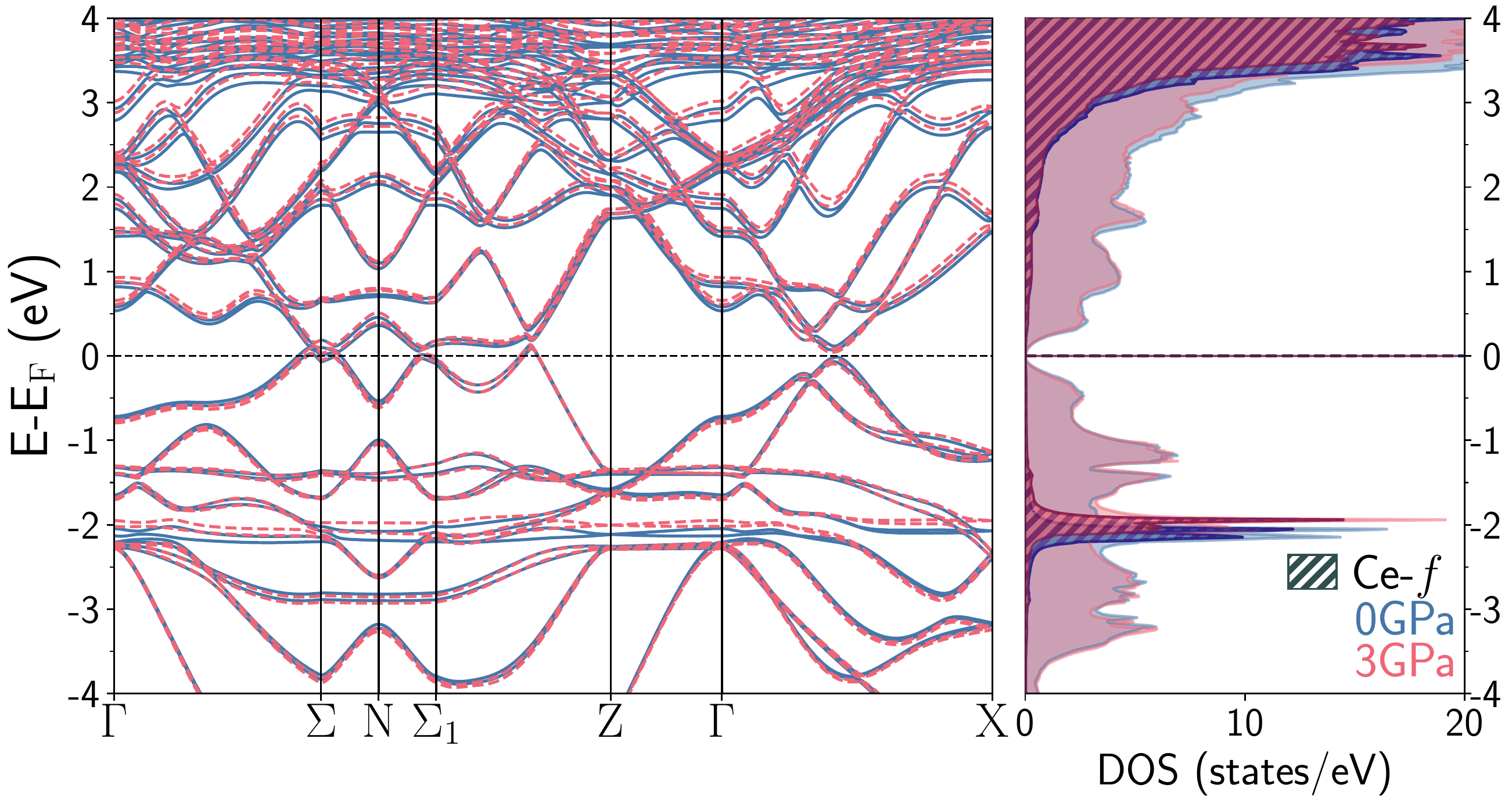}
	\caption{Electronic bands and DOS at ambient pressure (blue) and 3~GPa (red). The hatched region of right panel corresponds to the partial DOS associated with Ce $f$ states.}
	\label{DOS}
\end{figure*}

At ambient pressure, the FM ordering transition in CeAlSi is marked by a singular magnetization $M(T)$ and sharp drop in electrical resistivity as a function of temperature $\rho(T)$ at $T_{C} \approx 8$~K. Figure~\ref{chi_rho} shows the effect of external pressure on the magnetic phase  (see Appendix A for additional data). Application of pressure linearly enhances $T_C(p)$ with a slope of 0.62(2)~K/GPa, driving $T_{C}$ from 7.8~K at ambient pressure to 9.4~K at 2.7~GPa (values taken from the resistivity data). More important is our finding that, for different pressures, the in-plane magnetization curves $M(H)$ at 2~K as a function of the applied magnetic field along the $[100]$ direction lie on top of each other. This result indicates a negligible pressure effect on the non-collinear planar magnetic structure found at ambient pressure \cite{yang2021noncollinear}.

\subsection*{Electronic band structure}

Our DFT calculations at ambient pressure and 3~GPa, which incorporate spin-orbit coupling and non-collinear magnetic order, reveal only a negligible effect of pressure on the electronic band structure and the electronic density of states (DOS) at the Fermi energy [see Fig~\ref{DOS}] as well as on the ordered moments and their orientation. The bands contributing to the hole pockets at the Fermi surface (FS) barely display any variation of their intercepts of the Fermi energy in $\mathbf{k}$-space, suggesting a negligible variation of the FS area (see Appendix D for further information). The only noticeable modification of the electronic structure is a small shift of the bands associated with Ce $f$-electrons to higher energies with respect to the Fermi energy. As these bands lie about 2~eV below the Fermi level, they most likely do not directly contribute to the transport properties.

\subsection*{Quantum oscillations}

\begin{figure}[!t]
	\includegraphics[width=0.9\linewidth]{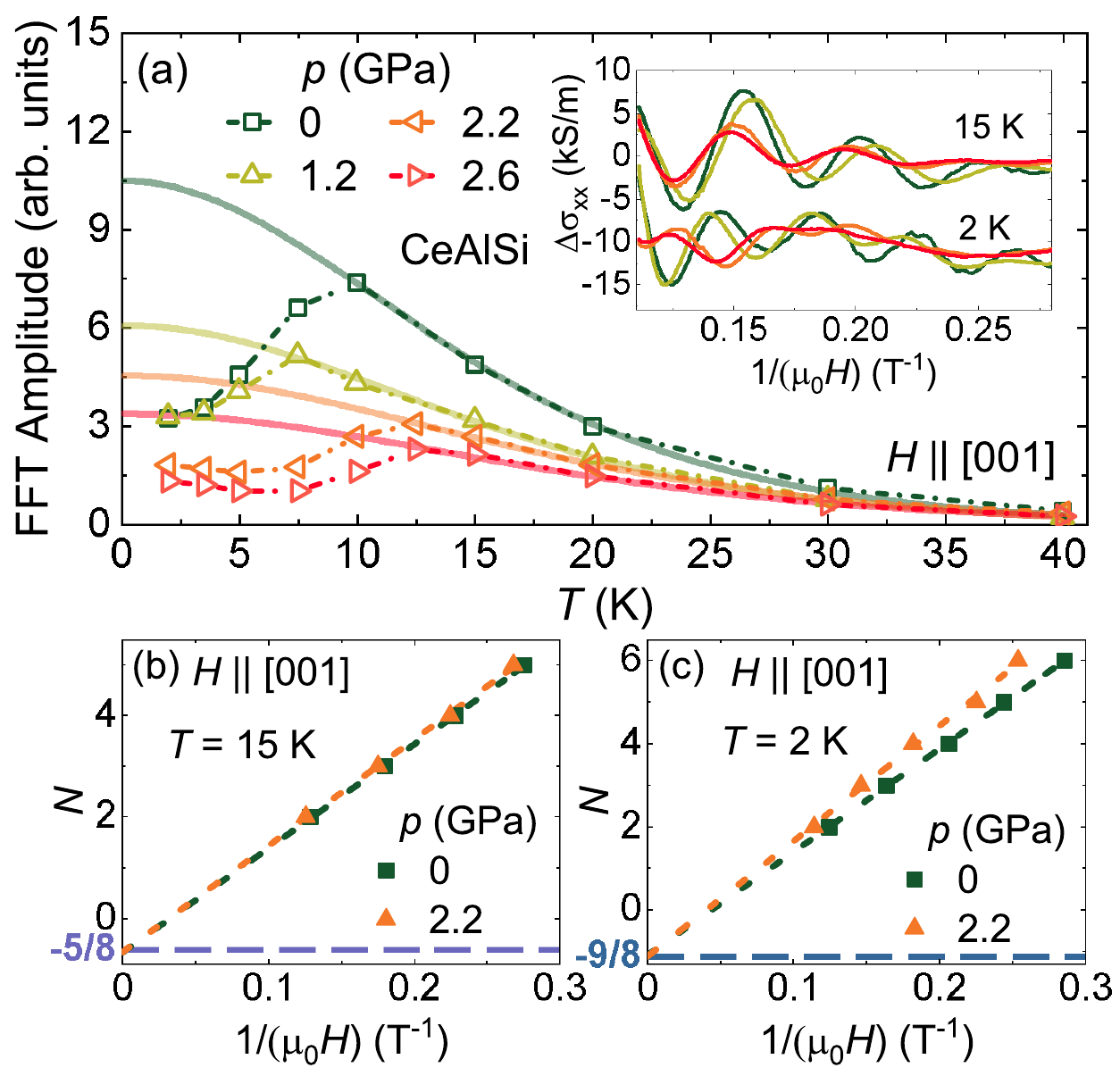}
	\caption{(a) Fast Fourier transformation (FFT) amplitude as a function of temperature at different applied pressures. The solid lines are simulations considering the best fits using the Lifshitz-Kosevich formula. The inset shows longitudinal conductivity for $H\parallel[001]$ after subtraction of a third order polynomial background $\Delta\sigma_{xx}$ as a function of $1/(\mu_{0}H)$ at 15~K (top) and 2~K (bottom) for selected pressures. The curves at 2~K were shifted by $-10$~kS/m for clarity. (b) and (c) Landau fan diagrams for CeAlSi at 15 and 2~K, respectively.}
	\label{DeltaSigma}
\end{figure}

Next we turn to the results of our QO measurements.
Longitudinal conductivity data $\sigma_{xx}$ well above $T_{C}$ at $T= 15$~K and in the FM state at $T= 2$~K at several pressures are presented in the inset of Fig.~\ref{DeltaSigma}(a), where we have subtracted a smooth background yielding $\Delta\sigma_{xx}$. Within the investigated field range, the $\Delta\sigma_{xx}$ data as well as its fast Fourier transform (FFT) analysis reveals a single QO frequency $f\approx 20(5)$~T, which is found to be independent of pressure and temperature (see Appendix B).
We notice two main features: i.\ the amplitude of the oscillations at 15~K is larger than that at 2~K and ii.\ the amplitude of the oscillations is suppressed by increasing pressure [see Fig.~\ref{DeltaSigma}(a)]. Generally, the thermal damping of the QO amplitude can be described by the Lifshitz-Kosevich (LK) formula \cite{shoenberg2009magnetic}. However, our FFT signal follows the LK prediction only in the paramagnetic (PM) region above $T_C$ [solid lines in Fig.~\ref{DeltaSigma}(a)]. In the FM state we observe an unusual reduction of the QO amplitude upon cooling. This remarkable response of the oscillation amplitude as a function of temperature has not been observed in any other members of the $Ln$Al$Pn$ family \cite{yang2021noncollinear,su2021multiple,xu2021shubnikov,gaudet2020incommensurate,wang2022temperature}, and was previously reported in just a few materials \cite{honold1997importance,wu2019anomalous}. In SmSb, for instance, a sudden decrease of the Shubnikov-de Haas oscillations takes place once the material becomes antiferromagnetic, which was conjectured to be due to the presence of a nontrivial Berry phase \cite{wu2019anomalous}.

To further analyze the QO, Landau fan diagrams are shown in Figs.~\ref{DeltaSigma}(b) and \ref{DeltaSigma}(c). Our analysis indicates a change in the nature of the topological properties between the PM and FM phase. At 15~K in the PM phase the intercept is around $-5/8$, which suggests the presence of topologically trivial charge carriers \cite{wang2016anomalous}. In contrast to that, at 2~K the intercept is $-9/8$, which for 3D magnetic WSMs can be associated with linear dispersive charge carriers and a nontrivial Berry phase \cite{wang2016anomalous}.

Our QO data suggest that the momentum space separation between nearby Weyl nodes with opposite topological charges gets enhanced in the FM state leading to a change in FS topology - from one which encloses both Weyl nodes above $T_C$ to a split FS enclosing isolated well-separated Weyl nodes below $T_C$. An enhanced separation of the Weyl nodes in the FM phase has been also previously found in band-structure calculations (SI of Ref.\ \cite{yang2021noncollinear}). This can explain the change in the intercept in our Landau fan plot. Such a change in the FS topology could 
nonetheless preserve the area of certain extremal orbits, so that the observed QO frequency can remain nearly unchanged 
(see Appendix E for details). If the topological Fermi pockets are only weakly 
split below $T_C$, 
the large density of states due to proximity 
to the Lifshitz transition \cite{fontana2021topological} can lead to an increased scattering rate for states on the extremal orbits, thus enhancing the 
Dingle temperature and suppressing the QO amplitude for $T <  T_C$. We note that our results cannot rule out other possible scenarios for the suppression of the amplitude of the quantum oscillations upon cooling. However, a Lifshitz transition in the FS of CeAlSi can explain both, the suppression of the QO and the phase shift upon entering the FM phase revealed by our measurements.

\begin{figure}[!t]
	\includegraphics[width=0.98\linewidth]{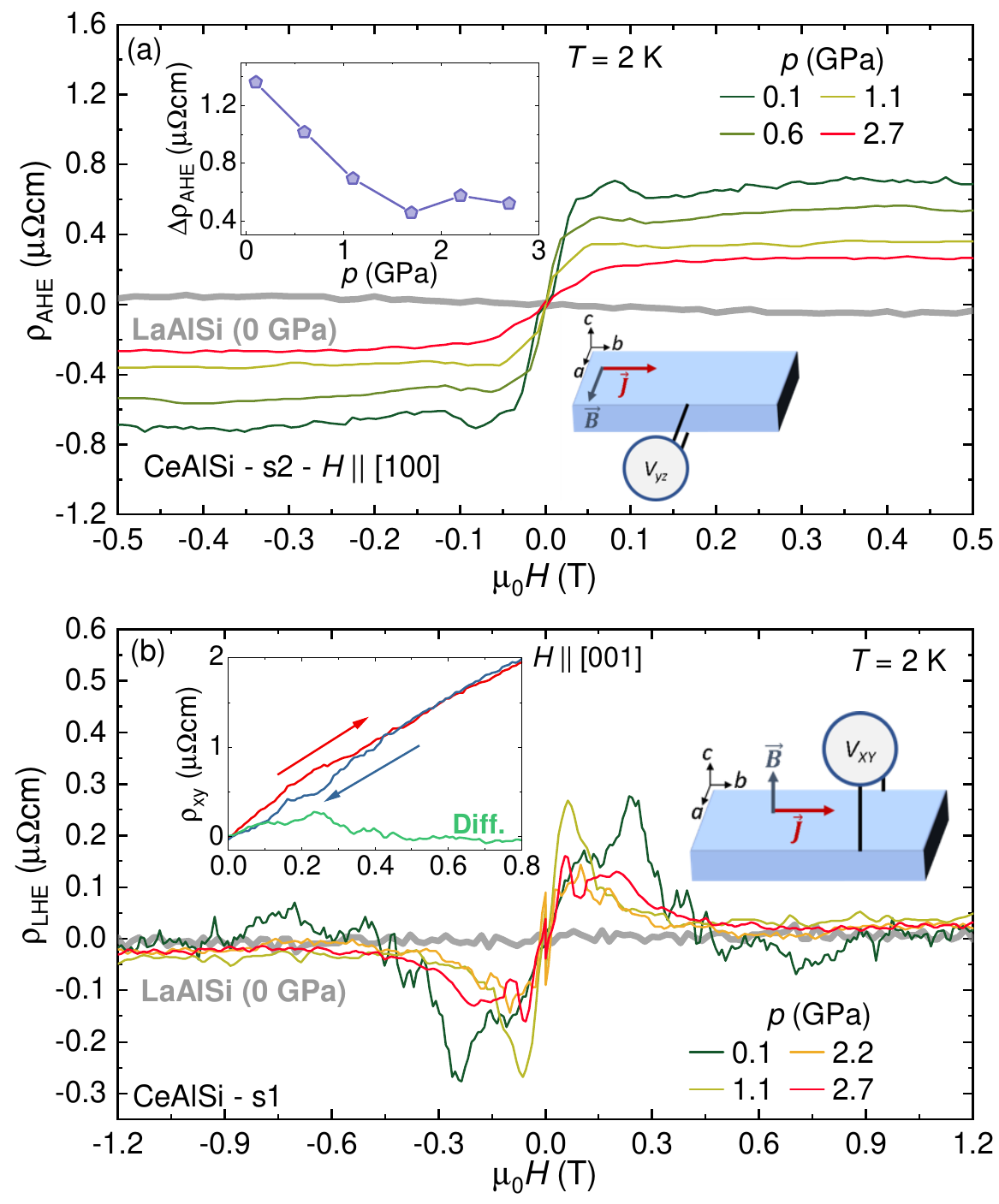}
	\caption{ (a) Anomalous Hall effect (AHE) at 2~K as a function of magnetic fields $H\parallel[001]$ for different applied pressures. The top inset shows the AHE jump as a function of pressure and the bottom inset displays a scheme of the circuit used in this measurement. s1 and s2 denote samples 1 and 2, respectively.
		(b) Loop Hall effect (LHE) at 2~K as a function of magnetic field for $H\parallel[001]$ for selected applied pressures. The left inset shows the Hall resistivity measured upon increasing (red) and decreasing (blue) magnetic field and the difference of both curves (green) at 0.1~GPa. The right inset displays a schematic drawing of the measurement circuit. Data of the nonmagnetic reference material LaSiAl at ambient pressure is shown as gray line in both panels}
	\label{AHE}
\end{figure}

\subsection*{Hall effect}

For magnetic field along the $[100]$ direction and current along $[010]$ [see sketch in Fig.~\ref{AHE}(a)] we find a large AHE in CeAlSi in its ferromagnetic state. The AHE signal has been extracted by fitting the Hall resistivity to the form $\rho_{yz}(H) = R_{0}H + \rho_{\rm AHE}$, where $R_{0}$ is the ordinary Hall effect coefficient and $\rho_{\rm AHE} = R_s M_x$ with $R_s$ being the anomalous Hall coefficient and $M_x$ being the magnetization along $[100]$ (see Appendix C).
We confirm that an AHE is absent in the non-magnetic analog LaAlSi  [Fig.~\ref{AHE}(a)]. We have fitted the longitudinal and ordinary Hall conductivities to a simple two-band model to obtain information on the density of the electron- and hole-like charge carriers and their mobilities (see Appendix C). At low temperatures and ambient pressure we find $5.9(1) \times 10^{19}$~holes/cm$^3$ and $2.5(1) \times 10^{19}$~electrons/cm$^{3}$. The application of external pressure suppresses the extracted hole density only slightly, which reaches $4.6(1) \times 10^{19}$~holes/cm$^3$ at 2~K and 2.6~GPa, whereas the electrons density remains nearly unchanged. Moreover, the corresponding mobilities at 2~K and ambient pressure are about $1.4(1) \times 10^3$~cm$^2$/Vs ($3.2(1) \times 10^3$~cm$^2$/Vs) for holes (electrons). These values are nearly unaffected by application of external pressure and are on the same order of magnitude compared with other Weyl semimetals \cite{liu2018giant,he2021large}.  Application of external pressure suppresses the jump of the AHE (defined as difference in $\rho_{\rm AHE}$ between positive and negative magnetic fields) up to 1.5~GPa [top inset of Fig.~\ref{AHE}(a)]. Above 1.5~GPa the anomalous Hall jump saturates to around 0.5(1)~$\mu \Omega$cm.
As we have shown above, the $M(H)$ curves taken at different pressures fall on top of each other [see Fig.~\ref{chi_rho}(b)], suggesting the absence of changes in the magnetic structure as a source for the suppression of the AHE. Moreover, the electronic bands close to the Fermi level are only slightly affected by pressure [see Fig.~\ref{chi_rho}(c)], making it unlikely that
this significant decrease results from a pressure-induced change in the position of the Weyl nodes. We find that while $R_s$ scales nearly linearly with $\rho_{xx}$ for 
pressure $\lesssim\! 1$\,GPa, the scaling deviates significantly from linear behavior at higher pressures (see Appendix C). A linear relation between $R_s$ and $\rho_{xx}$ suggests that the observed AHE at ambient pressure has a significant extrinsic skew-scattering contribution \cite{nagaosa2010anomalous}, which gets suppressed at high pressures. Given the robustness of the electronic structure and magnetic order against pressure, 
the most plausible explanation for this is a pressure-dependent change in the nature or distribution of domain walls.
Previous work has shown that Weyl fermions can undergo skew scattering from magnetic domain walls which contain the axis of the average magnetization, leading to an extrinsic contribution to the AHE qualitatively consistent with our data \cite{sorn2021domain}.

An even more unusual Hall response is observed for field applied along $[001]$ [see sketch in Fig.~\ref{AHE}(b)]. We note that in this geometry the magnetic field is applied perpendicular to the ferromagnetically ordered moments in the $\left<001\right>$ plane and this Hall response thus cannot arise from the bulk in-plane magnetization. This so-called loop Hall effect (LHE) is displayed in Fig.~\ref{AHE}(b). It is only observed in the ferromagnetic regime, displays hysteresis even in the absence of any observable $M_z$ magnetization hysteresis, and is absent in the non-magnetic analog LaAlSi. $\rho_{\rm LHE}(H)$ is obtained by recording the Hall resistivity $\rho_{xy}$ for $H\parallel[001]$ upon increasing and decreasing magnetic field and taking the difference between both curves, as shown in the left inset of Fig.~\ref{AHE}(b) for 0.1~GPa as an example. Similar to the AHE, the application of external pressure leads to a decrease in the LHE [see Fig.~\ref{AHE}(b)].
While the existence of the LHE in CeAlSi has been argued to be tied to the presence of the Weyl nodes near the Fermi energy \cite{yang2021noncollinear}, no clear
physical mechanism has been provided for its origin.

\section{DISCUSSION}\label{discussion}

\begin{figure}[!tb]
	\includegraphics[width=1.0\linewidth]{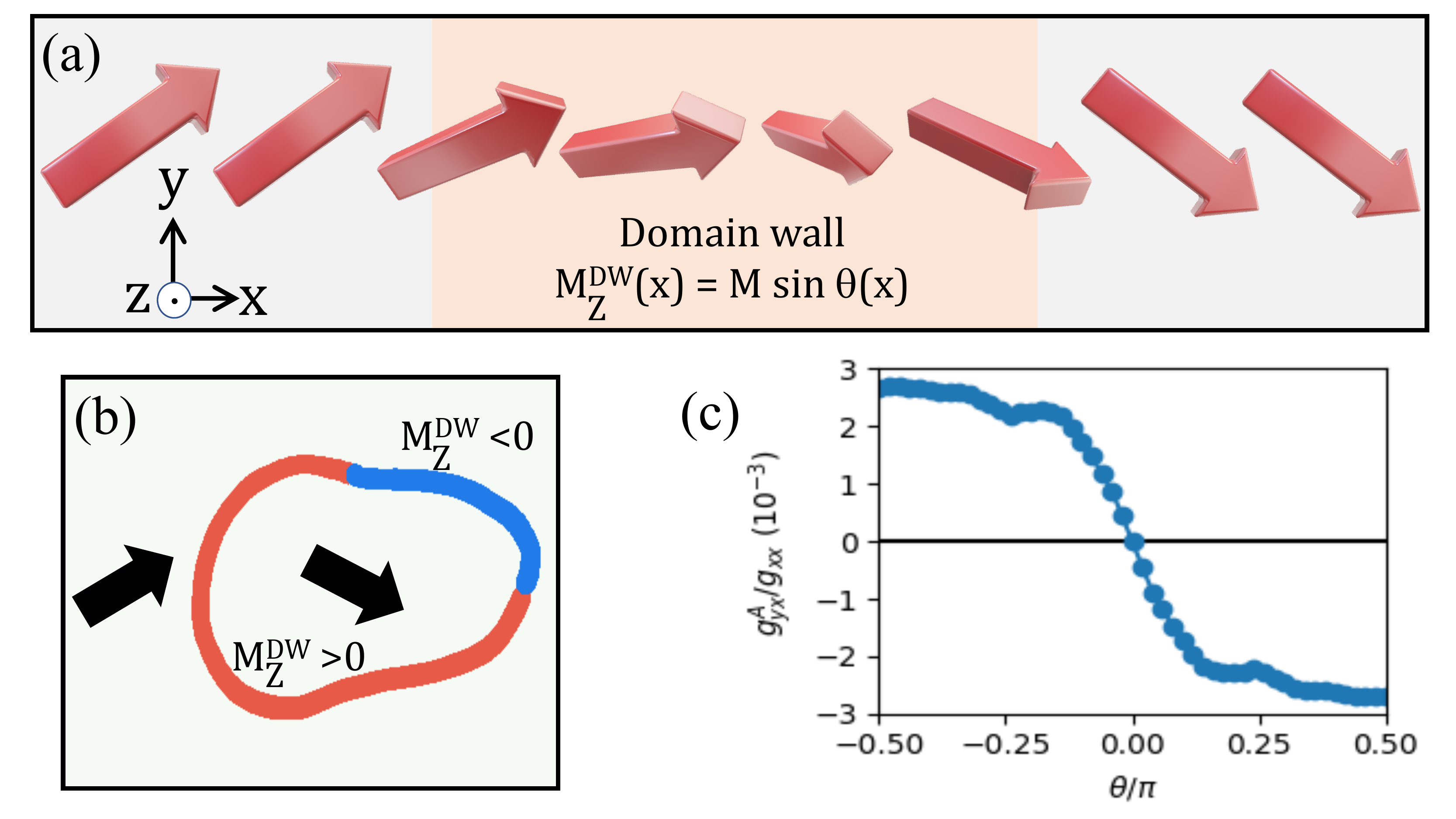}
	\caption{(a) Domain wall between two bulk magnetic
	domains showing twisted magnetization configuration 
	with $M^{DW}_z(x)=M \sin\theta(x)$. (b) Skew scattering of Weyl fermions off hysteretic domain wall loops, with red regions having $M^{DW}_z > 0$ and blue having $M^{DW}_z < 0$, provides a mechanism for the loop Hall effect (LHE). The black arrows are classical representations of the Weyl fermions trajectories. (c) Calculated
	LHE shown as a ratio of Landauer conductances,
	versus the maximal out of plane tilt angle $\theta$ of the domain wall magnetization.}
	\label{LHE_WeylScatt}
\end{figure}

In the following we present a simplified model for Weyl fermions in a non-centrosymmetric FM, and show that a domain wall scattering mechanism, similiar to that previously explored to understand the AHE \cite{sorn2021domain}, can also lead to the LHE for domain walls which are perpendicular to the average magnetization. Our key idea here is that the bulk magnetic domains in CeAlSi host an in-plane magnetization with hard axis along $[001]$, so that the out-of-plane {\it bulk} contribution to the field-induced magnetization $M^{\rm bulk}_z$ is not expected to be hysteretic as we tune the magnetic field $H_z$. We instead argue that the hysteretic LHE must be attributed to the hysteretic {\it domain wall} magnetization $M^{\rm DW}_z$ as we tune $H_z$, as schematically depicted in Fig.~\ref{AHE}(a) and \ref{AHE}(b). Our calculations show that the intra-node skew scattering of Weyl fermions as they cross a domain wall with nonzero $M^{\rm DW}_z$ can explain the LHE.

To illustrate this physics, we study a model with $4$ pairs of Weyl nodes in the $k_z=0$ plane (see Appendix E for details). These could be viewed as a caricature of the $W_3'$ Weyl nodes found to lie $\sim 46$~meV above the Fermi level close to the $k_z=0$ plane in CeAlSi \cite{sakhya2022observation,yang2021noncollinear}. For a single Weyl node, with chirality $+1$, we consider a simple linearized Hamiltonian:
\begin{eqnarray}
	\mathcal{H}_+ &=& v_F \sigma_i G^{(+)}_{ij} q_j + \sigma_i M_i
\end{eqnarray}
where $v_F$ is the nodal velocity, $\sigma$ is the spin Pauli matrix, $q$ denotes the momentum relative to the Weyl node position, and the tensor $G_{ij}$ is chosen to yield an elliptical Fermi surface at fixed $k_z$ with its major axis rotated away from the $x,y$ axes. The Hamiltonian for the other Weyl nodes can be reconstructed using symmetries. The Weiss field ${\bf M}$ is nonzero in the magnetically ordered phase and tunes the momentum of the Weyl node; we assume this leads to topologically nontrivial FS pockets enclosing single Weyl nodes.
For a magnetic field $H_z$ applied along the $z$-axis, the system will support domains of ${\bf M}$, with magnetization aligned along different in-plane directions, which are separated by domain walls. As shown in \cite{sun2021mapping}, the domain wall magnetization in such cases supports an out-of-plane component $M^{\rm DW}_z$. Fig.~\ref{LHE_WeylScatt}(c) shows the computed Hall conductance scaled by the longitudinal conductance showing that it is an odd function of $M^{\rm DW}_z$ (see Appendix F for details). Crudely, this small Landauer conductance \cite{datta1997electronic,nazarov2009quantum} ratio is expected to be related to the ratio of loop Hall to longitudinal resistivity - our experiments show that $\rho_{xy}^{\rm LHE}/\rho_{xx} \!\sim\! 10^{-3}$, in reasonable agreement with the theoretical estimate in Fig.~\ref{LHE_WeylScatt}. We thus propose that the mechanism for the aptly named LHE is the skew scattering of Weyl fermions off hysteretic domain wall loops or surfaces. Since we expect the domain wall magnetization $M^{\rm DW}_z \!\ll\! M^{\rm bulk}_z$, the hysteretic behavior of $M^{\rm DW}_{z}$ cannot be resolved in bulk magnetization measurements. Our model might also help to understand the observation of a similar LHE reported previously in other compounds, in which Weyl fermions were predict to exist \cite{ma2015mobile,ueda2018spontaneous, yamaji2014metallic}.

\section{CONCLUSIONS}

In summary, our study emphasizes, through a key tuning parameter (hydrostatic pressure), the importance of ferromagnetism for the low temperature topological features in CeAlSi. Our QO data show a difference in the Berry phase above and below $T_C$, indicating that FM ordering shifts oppositely charged Weyl nodes away from each other in momentum space, leading to a change in the Fermi-surface topology. We have argued that this also leads to an increase in the scattering rate below $T_C$, and thus to a drop in the amplitude of Shubnikov -- de Haas oscillations in contrast to the conventional LK formula. This result calls for angular dependent Shubnikov -- de Haas and de Haas -- van Alphen experiments in an extended field range.
We have also discovered pressure dependent changes in the AHE and LHE below $T_C$. Since our DFT calculations indicate that the electronic band structure is robust against pressure, we argue that these changes in the AHE and LHE must arise from differences in domain wall defects and we have shown how Weyl fermions scattering off hysteretic domain walls can lead to the LHE. 

\section*{DATA AVAILABILITY}
Data that underpin the ﬁndings of this study are available at Edmond – the open research data repository of the Max Planck Society at \cite{EDMOND}.

\begin{acknowledgments}
	
We acknowledge fruitful discussions with A.\ P.\ Mackenzie. We also thank U.\ Burkhardt for carrying out energy dispersive x-ray analysis on the samples. This project has received funding from the European Union’s Horizon 2020 research and innovation programme under the Marie Sk\l{}odowska-Curie grant agreement No 101019024. This work was also supported by the S\~ao Paulo Research Foundation (FAPESP) grants 2017/10581-1, 2018/11364-7, 2020/12283-0, CNPq grants $\#$ 304496/2017-0, 310373/2019-0 and CAPES, Brazil. This research was financially supported by the Natural Sciences and Engineering Research Council of Canada (NSERC), under the Discovery Grants program grant No. RGPIN-2016-06666. Computations were made on the supercomputers Beluga and Narval managed by Calcul Québec and the Digital Research Alliance of Canada. The operation of these supercomputers is funded by the
Canada Foundation for Innovation, the Minist\`ere de la Science, de l'\'Economie et de l'Innovation du Qu\'ebec, and the Fonds de recherche du Qu\'ebec – Nature et technologies. V.B.-C. and M.C are members of the Regroupement qu\'eb\'ecois sur les mat\'eriaux de pointe (RQMP).

\end{acknowledgments}

\section*{APPENDICES} 

\section*{ APPENDIX A: Electrical resistivity}

 \begin{figure*}[!tbh]
	\includegraphics[width=0.9\linewidth]{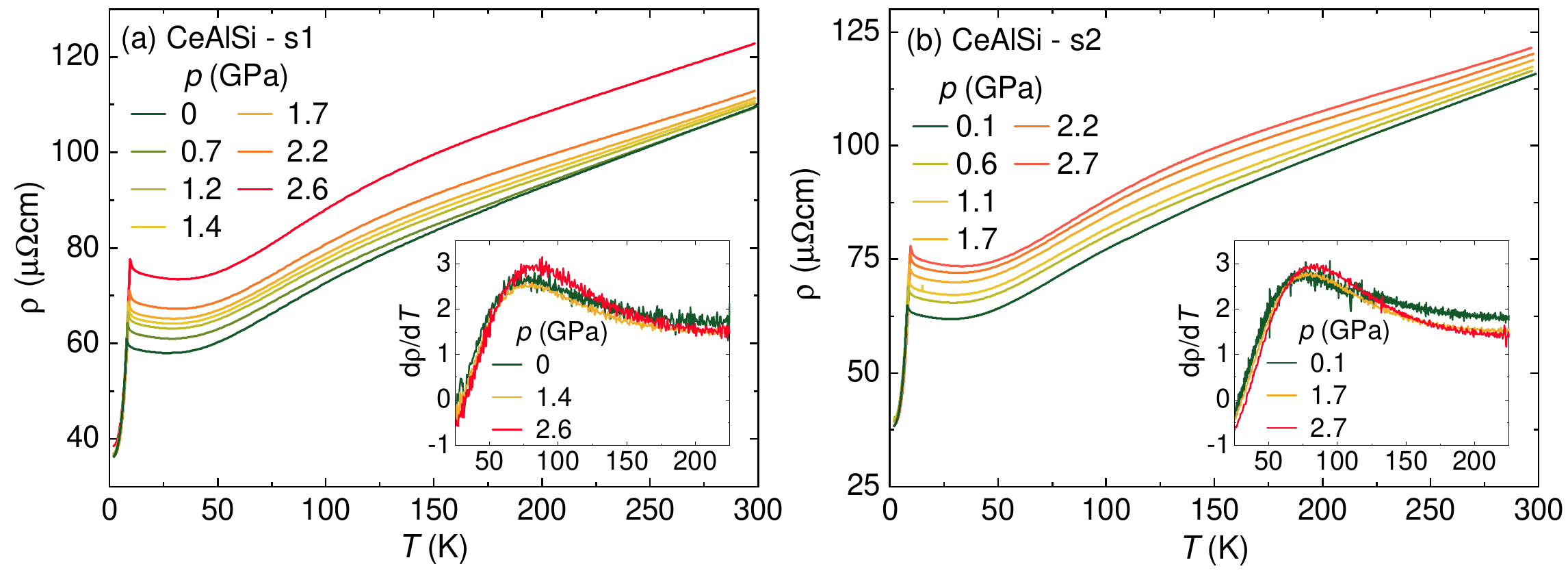}
	\caption{(a) and (b) Electrical resistivity ($\rho$) as a function of temperature at several pressures for two different samples of CeAlSi. The insets displays a magnified view of the low-temperature range. The insets show the temperature derivative of $\rho$ as a function of temperature at several pressures for two different samples of CeAlSi.}
	\label{RxT}
\end{figure*}

Figures~\ref{RxT}(a) and \ref{RxT}(b) present the electrical resistivity ($\rho$) as a function of temperature at several pressures for two different samples of CeAlSi. At high temperatures $\rho(T)$ exhibits a metallic behavior for both samples at all studied pressures. Moreover, a clear kink is observed at low-temperatures characterizing the ferromagnetic transition, in good agreement with previous reports at ambient pressure \cite{yang2021noncollinear,xu2021picoscale,sun2021mapping}. A broad shoulder is observed at around 80~K. It shows up as a maximum in the temperature derivative of the electrical resistivity [see insets of Figs.~\ref{RxT}(a) and \ref{RxT}(b)]. The shape and the position of the maximum is nearly unaffected by the application of external pressure, suggesting that the gap between the ground state and the first excited crystalline electrical field state does not change with increasing pressure.

\section*{APPENDIX B: Quantum Oscillations}

 \begin{figure*}[!tbh]
	\includegraphics[width=0.8\linewidth]{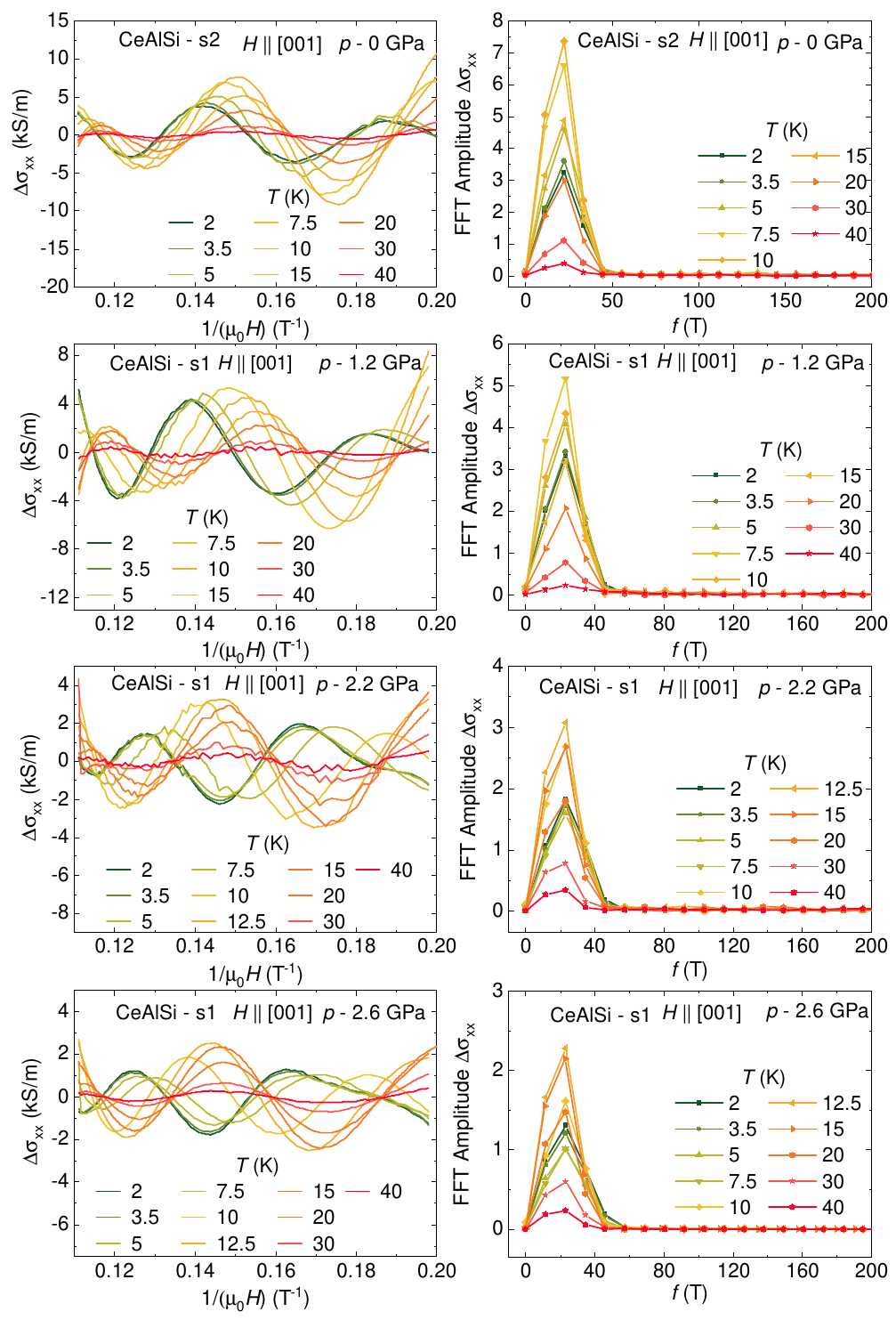}
	\caption{(Left panels) Longitudinal conductivity after subtraction of a third order polynomial $\Delta\sigma_{xx}$ as a function of $1/(\mu_{0}H)$ at several temperatures for selected pressures. (Right panels) Fast Fourier transformation of $\Delta\sigma_{xx}$ shown in the left.\vspace{2cm}}
	\label{QO}
\end{figure*}

The left panels of Fig.~\ref{QO} present the longitudinal conductivity measured with $H\parallel[001]$ after subtraction of a third order polynomial (fit between 5 and 9~T)  $\Delta\sigma_{xx}$ as a function of $1/(\mu_{0}H)$. We note that quantum oscillations are clearly seen up to 40~K at all studied pressures. Furthermore, the unusual behavior of the quantum oscillations amplitudes can be seen by the naked eye. The oscillations in the paramagnetic state at 15~K are more pronounced than the oscillations in the ferromagnetic state at 2~K. The panels on the right side of Fig.~\ref{QO} present the Fast Fourier transformation (FFT) of $\Delta\sigma_{xx}$, using a Hamming window from 0.2 to 0.11~T$^{-1}$, as a function of frequency ($f$) at several temperatures for selected pressures. Only one oscillation frequency $f\approx20(5)$~T is present. It is unaffected by changes in pressure and/or temperature.


\begin{figure}[!tbh]
	\includegraphics[width=0.85\linewidth]{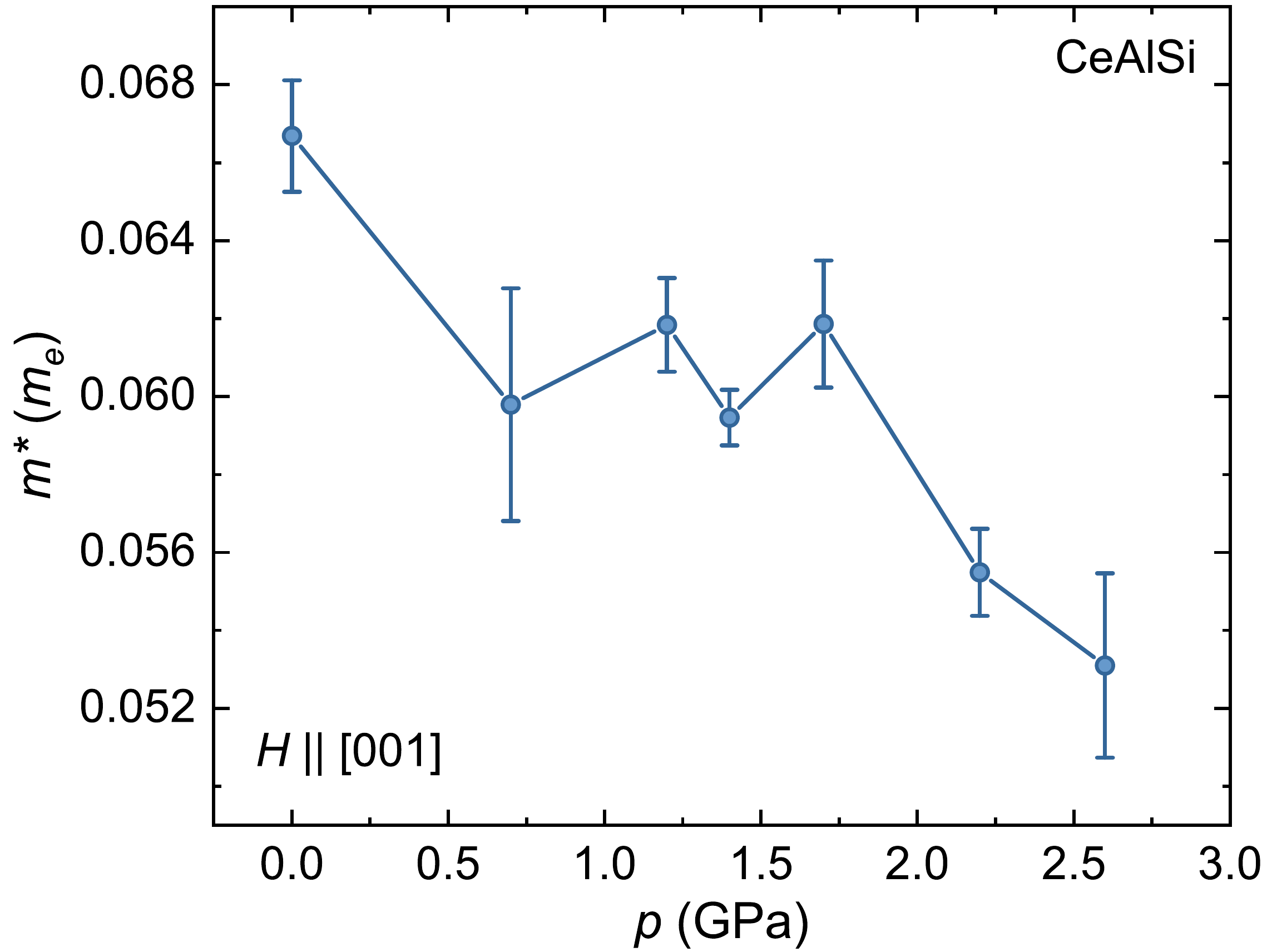}
	\caption{Effective mass $m^{*}$ as a function of pressure for magnetic fields parallel to $[001]$.}
	\label{effmass}
\end{figure}

 The effective mass ($m^{*}$) was estimated in the paramagnetic state of CeAlSi by fitting the FFT amplitude as a function of temperature by the Lifshitz-Kosevich (LK) formula \cite{shoenberg2009magnetic}:
\begin{equation}
	R_{T} = \frac{\alpha T m^{*}}{B \sinh (\alpha T m^{*}/B)},
\end{equation}
in which $\alpha=2 \pi^{2} k_{B}/e \hbar \approx 14.69$~T/K, $T$ is the temperature, $B$ is the magnetic field and $m^{*}$ the effective mass. As shown in Fig.~\ref{effmass}, the application of external pressure leads to a decrease in the value of $m^{*}$.

\section*{APPENDIX C: Hall Effect}

\subsection*{Anomalous Hall Effect}

\begin{figure}[!tbh]
	\includegraphics[width=0.85\linewidth]{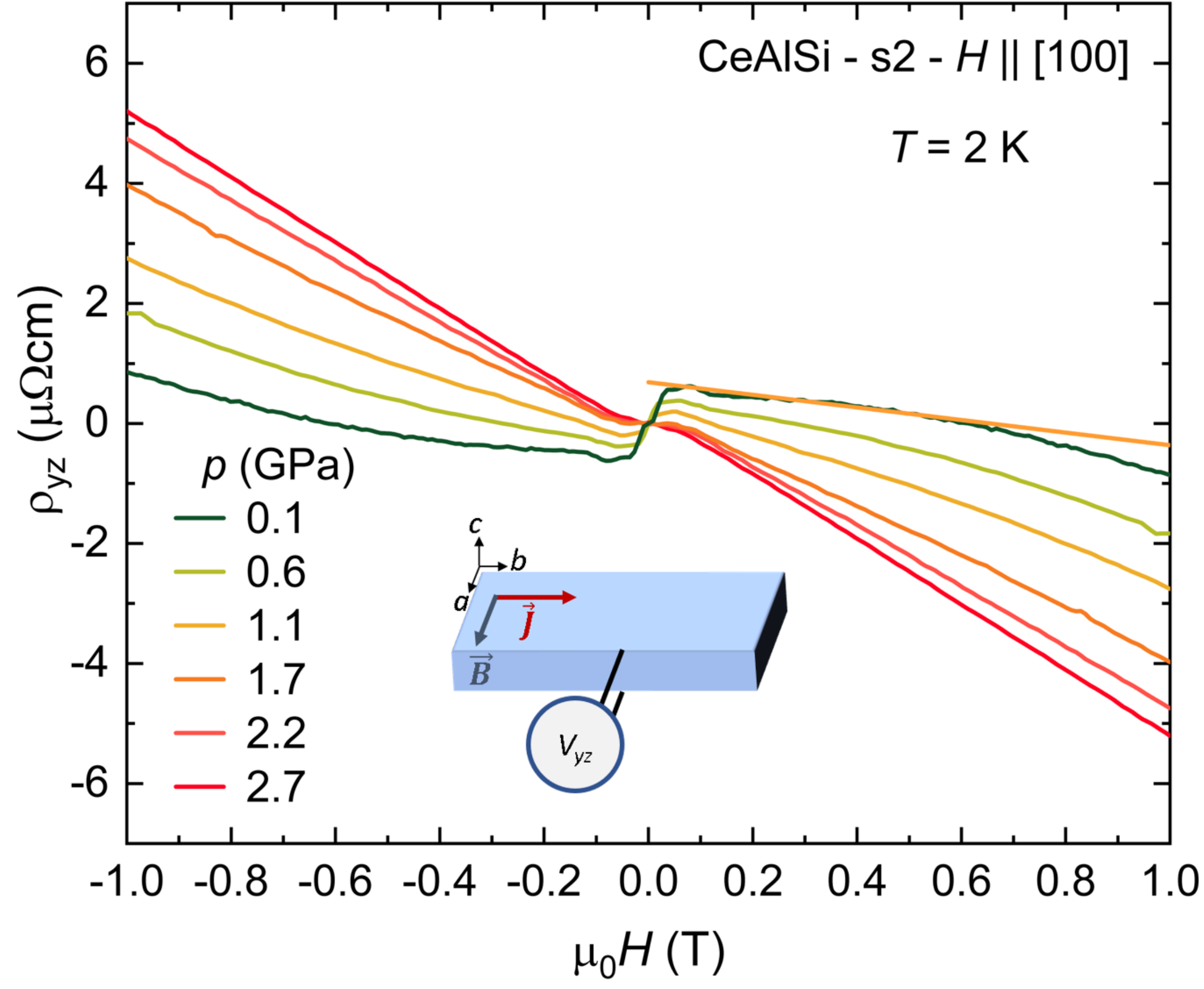}
	\caption{Hall resistivity ($\rho_{yz}$) as a function of magnetic field applied parallel to [100] at 2~K for several pressures. The solid orange line is an extrapolation of a linear fit performed in the range 0.2~T~$\leqslant H \leqslant $~0.6~T, which yields the ordinary background of $\rho_{yz}$.}
	\label{AHE_backgr}
\end{figure}

Figure~\ref{AHE_backgr} presents the Hall resistivity ($\rho_{yz}$) as a function of magnetic field at 2~K for several pressures. A linear background was determined by performing a linear fit in the range 0.2~T~$\leqslant H \leqslant $~0.6~T. We obtained the anomalous Hall ($\rho_{\rm AHE}$) effect by subtracting the linear background using $\rho_{yz} = R_{0}H + \rho_{\rm AHE}$.

\begin{figure}[!tbh]
	\includegraphics[width=0.85\linewidth]{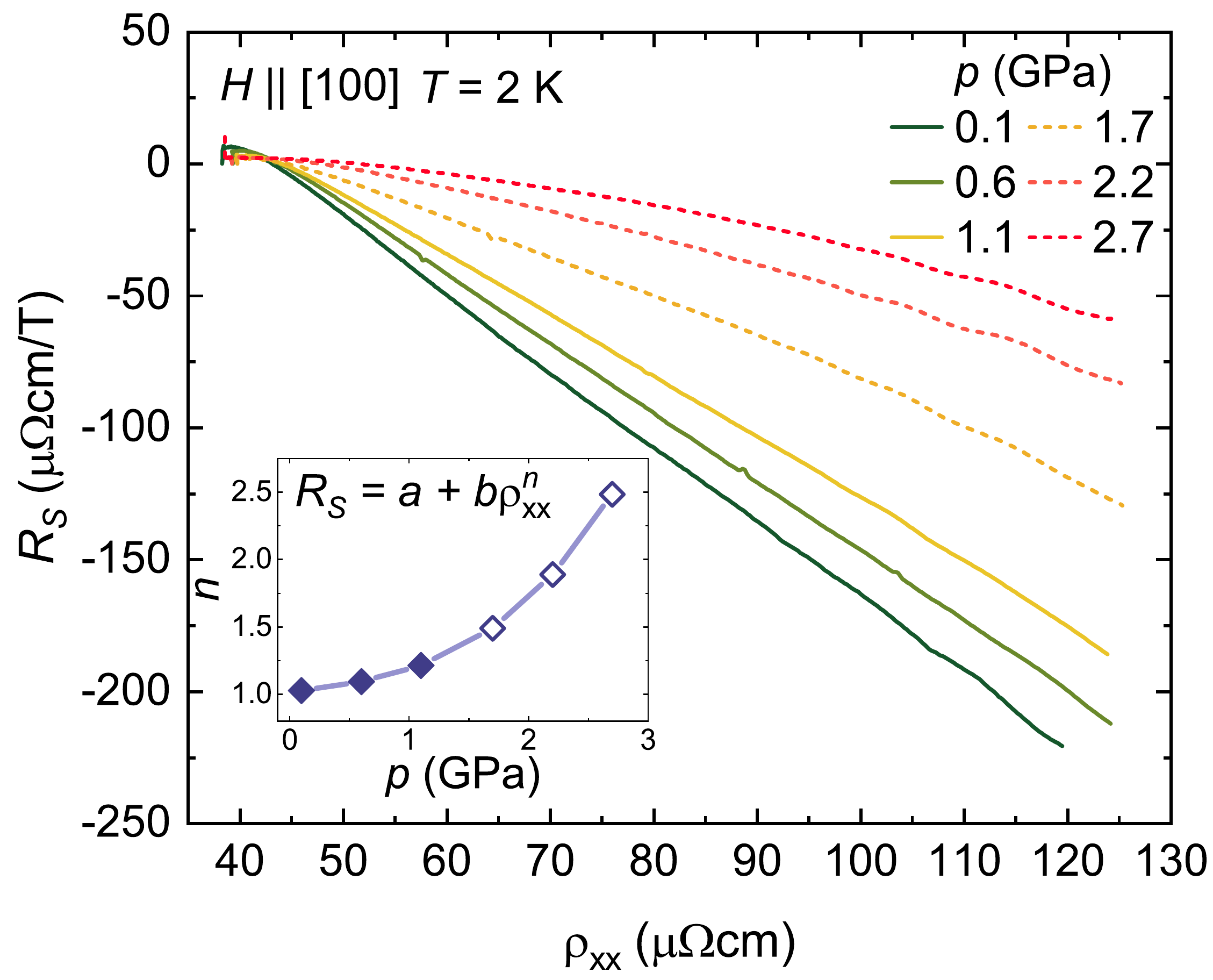}
	\caption{Anomalous Hall coefficient ($R_{S}$) as a function of the longitudinal resistivity ($\rho_{xx}$) at several pressures.}
	\label{Rs_rhoxx}
\end{figure}

As we can see in Fig.~\ref{Rs_rhoxx}, the linear dependence between the anomalous Hall coefficient ($R_{S}$)  and the longitudinal resistivity ($\rho_{xx}$) characterizes the presence of a skew scattering contribution to the AHE in CeAlSi at ambient pressure \cite{nagaosa2010anomalous}. The observation of this contribution in a good metal regime can be attributed to domain wall scattering of Weyl fermions (see Sec.~\ref{discussion}), as this contribution should be the dominant one in highly conducting samples ($\sigma_{xx}  \geqslant 0.5 \times 10^6 $~($\Omega$cm)$^{-1}$) \cite{nagaosa2010anomalous}. Furthermore, the application of external pressure suppresses the linear relation between $R_{s}$ and $\rho_{xx}$, which is better seen in the inset of Fig.~\ref{Rs_rhoxx}, where the exponents obtained with allometric fits ($R_{S} = a + b \rho_{xx}^n$) are shown as a function of pressure. One can clearly see the increase of the exponent $n$  as a function of increasing pressure, reaching 1.21(1) at 1.1 GPa, indicating that the skew scattering contribution of the AHE from the domain walls is being suppressed by application of pressure.  The domains themselves (bulk) also contribute to the AHE. It is possible to differentiate both contributions, as analyzed in great detail in Ref.~\cite{sorn2021domain}, by considering that the domain wall scattering contribution to the AHE is limited by the electron mean free path, whereas the bulk contribution is not. The total Hall resistivity is therefore an average between the bulk and domain wall contributions. Our results suggests that at low pressures ($p \leqslant 1.5$~GPa) the AHE is dominated by the skew scattering contribution coming from the domain walls, while in the high-pressure range ($p \geqslant 1.5$ GPa), where the AHE is not skew scattering type, it is dominated by the contribution of the domains.

\subsection*{Two-band model fits}

\begin{figure*}[!tbh]
	\includegraphics[width=0.9\linewidth]{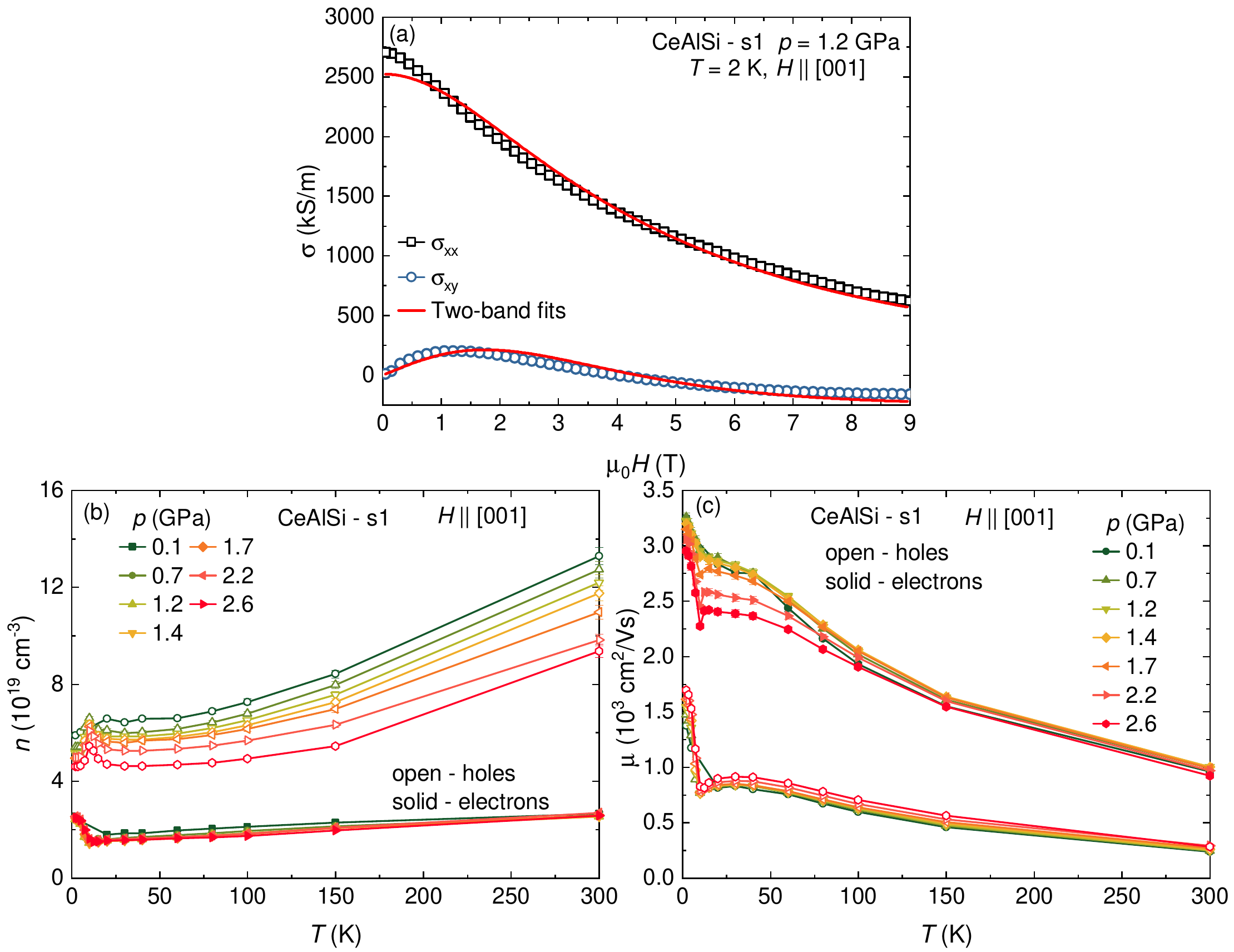}
	\caption{(a) Longitudinal ($\sigma_{xx}$) and Hall ($\sigma_{xy}$) conductivities at 2~K and 1.2~GPa. Carrier densities (b) and mobilities (c) obtained from the two-band fits as a function of temperature at several pressures.}
	\label{twobands}
\end{figure*}
To accurately estimate the carrier densities and mobilities of CeAlSi, we have simultaneously fit the longitudinal ($\sigma_{xx}$) and the Hall ($\sigma_{xy}$) conductivities considering a two-band model described by:  

\begin{eqnarray}\label{2band}
	\nonumber \sigma_{xx} &=& e\left(\frac{n_{e}\mu_{e}}{1 + \mu_{e}^{2}\left( \mu_{0}H\right)^2} + \frac{n_{h}\mu_{h}}{1 + \mu_{h}^{2}\left( \mu_{0}H\right)^2} \right) \\
		\nonumber \sigma_{xy} &=& e\left( \mu_{0}H\right)\left(\frac{n_{e}\mu_{e}^2}{1 + \mu_{e}^{2}\left( \mu_{0}H\right)^2} - \frac{n_{h}\mu_{h}^2}{1 + \mu_{h}^{2}\left( \mu_{0}H\right)^2} \right),
\end{eqnarray}

\noindent where $n$ denotes the electron ($e$) and hole ($h$) carrier densities, and $\mu_{e}$ and $\mu_{h}$ are the electron and hole mobilities, respectively. Figure~\ref{twobands}(a) presents a representative plot of the fits at 2~K and 1.2~GPa, in which a good agreement between the experimental data and the fits is observed. Figure~\ref{twobands}(b) displays the carrier densities as a function of temperature at several pressures. 
Figure~\ref{twobands}(c) shows the mobilities as a function of temperature at several pressures.

\section*{APPENDIX D: Bandstructure calculations}

\begin{figure*}[h]
	\centering
	\includegraphics[width=0.7\linewidth]{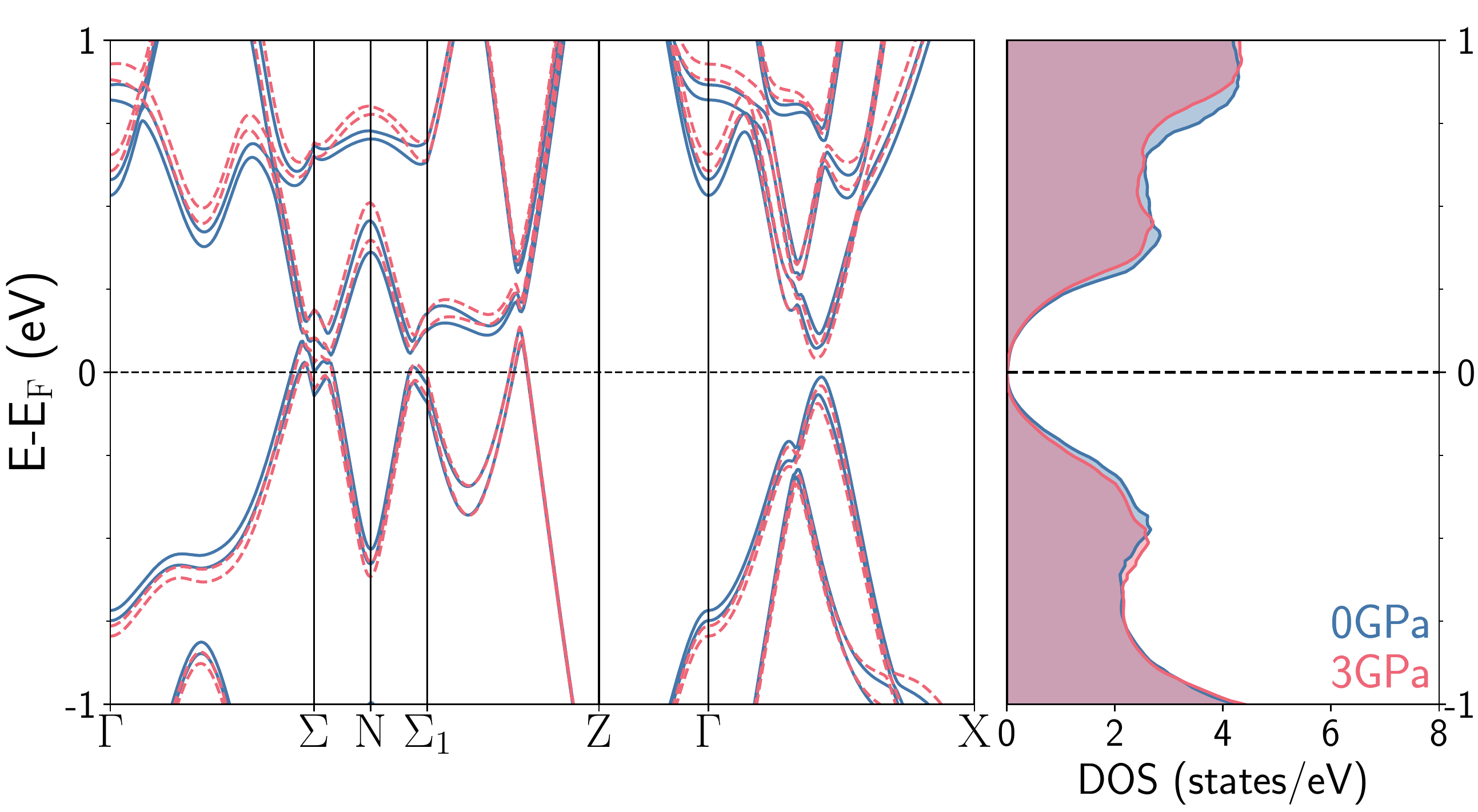}
	\caption{Electronic bands and DOS at ambient pressure (blue) and 3~GPa (red), zoomed in the vicinity of the Fermi level.}
	\label{fig:bstr_zoom}
\end{figure*}
\begin{figure*}[h]
	\centering
	\includegraphics[width=0.98\linewidth]{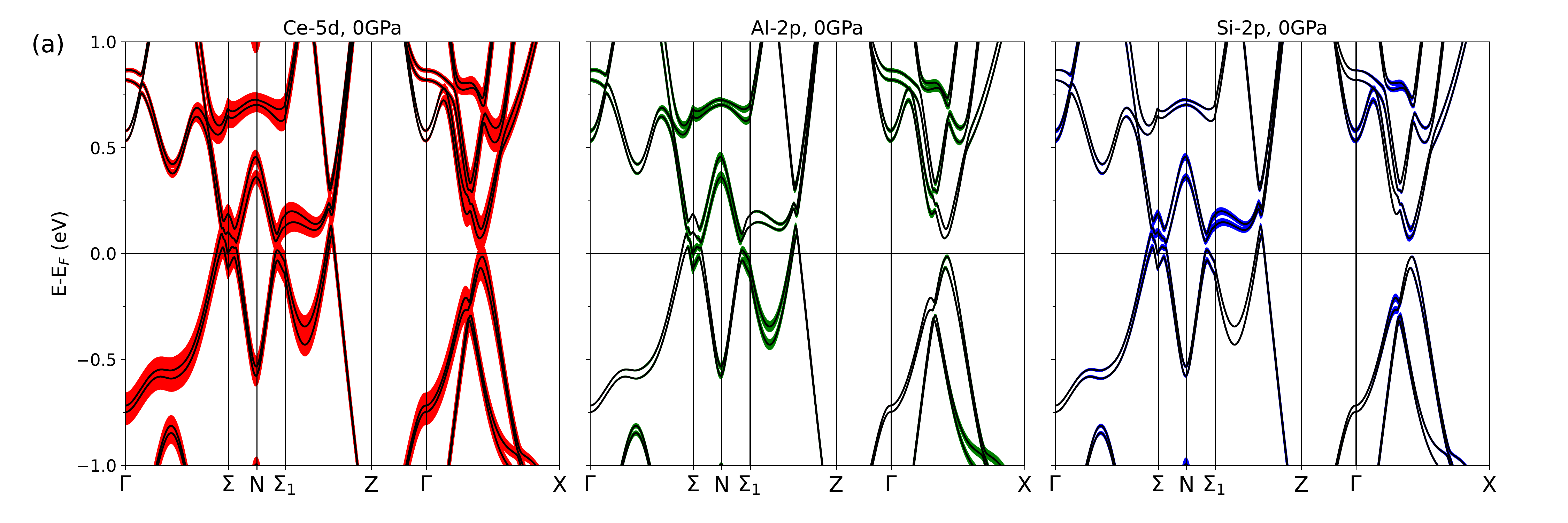}
	\includegraphics[width=0.98\linewidth]{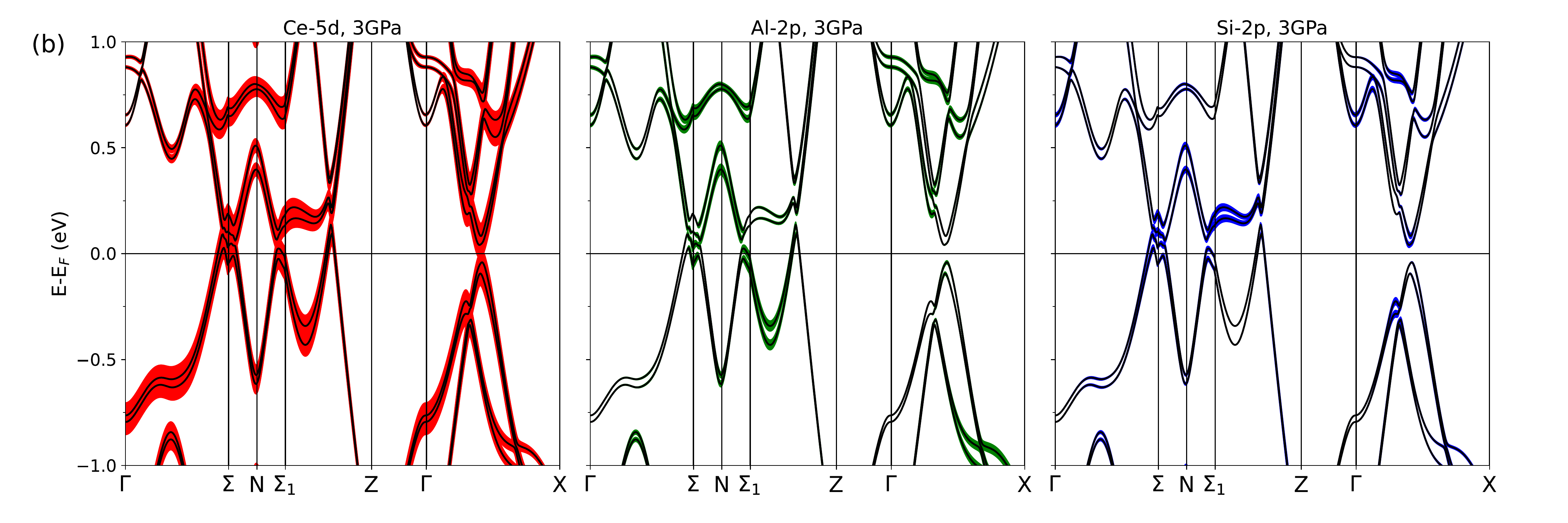}
	\caption{Orbital decomposition of the electronic wavefunction at (a) ambient pressure and (b) 3~GPa, zoomed in the vicinity of the Fermi level.}
	\label{fig:fatbands}
\end{figure*}

Figure~\ref{fig:bstr_zoom} shows the electronic bands and DOS zoomed in the vicinity of the Fermi level, to emphasize the negligible effect of pressure on the bands contributing to the AHE and LHE. Note, also that no crossing feature nor electron pocket was found in our ambient pressure calculation along the $\Gamma-X$ high symmetry path, in contrary to Fig.~3(a) of \cite{yang2021noncollinear}. This discrepancy could be attributed to the different exchange-correlation functional or to our use of theoretically relaxed lattice parameters, while \cite{yang2021noncollinear} used experimental values which, in the case of the PBE-GGA functional used in their paper, will be smaller than the theoretical one. Nevertheless, from the pressure dependence of the electronic bands relative to the Fermi level, an electron pocket could likely appear along this path upon further increasing the pressure. 

We further refine the analysis of the electronic structure by calculating the orbital decomposition of the electronic wavefunction inside the atom-centered PAW spheres for Ce $5d$ states (left panels, red), as well as for Al (middle panels, green) and Si (right panels, blue) $2p$ states, in the same energy range as Fig.~\ref{fig:bstr_zoom}. The relative weights of the different orbitals at ambient pressure (top panels) and~$3$~GPa (bottom panels) are essentially identical, thus confirming the negligible effect of pressure on the electronic bands.

The calculated magnetic structure does not display any significant differences between $0$~and~$3$~GPa either, in agreement with the experimental observations (see Fig.~1(b) of the main text). For $0$~GPa ($3$~GPa), we find a magnetic moment of 0.880 $\mu_B$ (0.878 $\mu_B$) inside the PAW spheres of the 2 inequivalent Ce atoms in the unit cell. Considering that one moment points mostly in the $\hat{x}$ direction and the other in the $\hat{y}$ direction with an angle of 87.3$^\circ$ (89.8$^\circ$) between them, we find a total net magnetic moment of 1.311 $\mu_B$ (1.284 $\mu_B$) oriented along $[110]$ for the whole unit cell. Note that the net size of the magnetic moment depends strongly on the choice of $U$.

\section*{APPENDIX E: SIMPLIFIED MODEL FOR WEYL NODES}

\begin{figure}[b]
\centering
\includegraphics[width = 0.98\columnwidth]{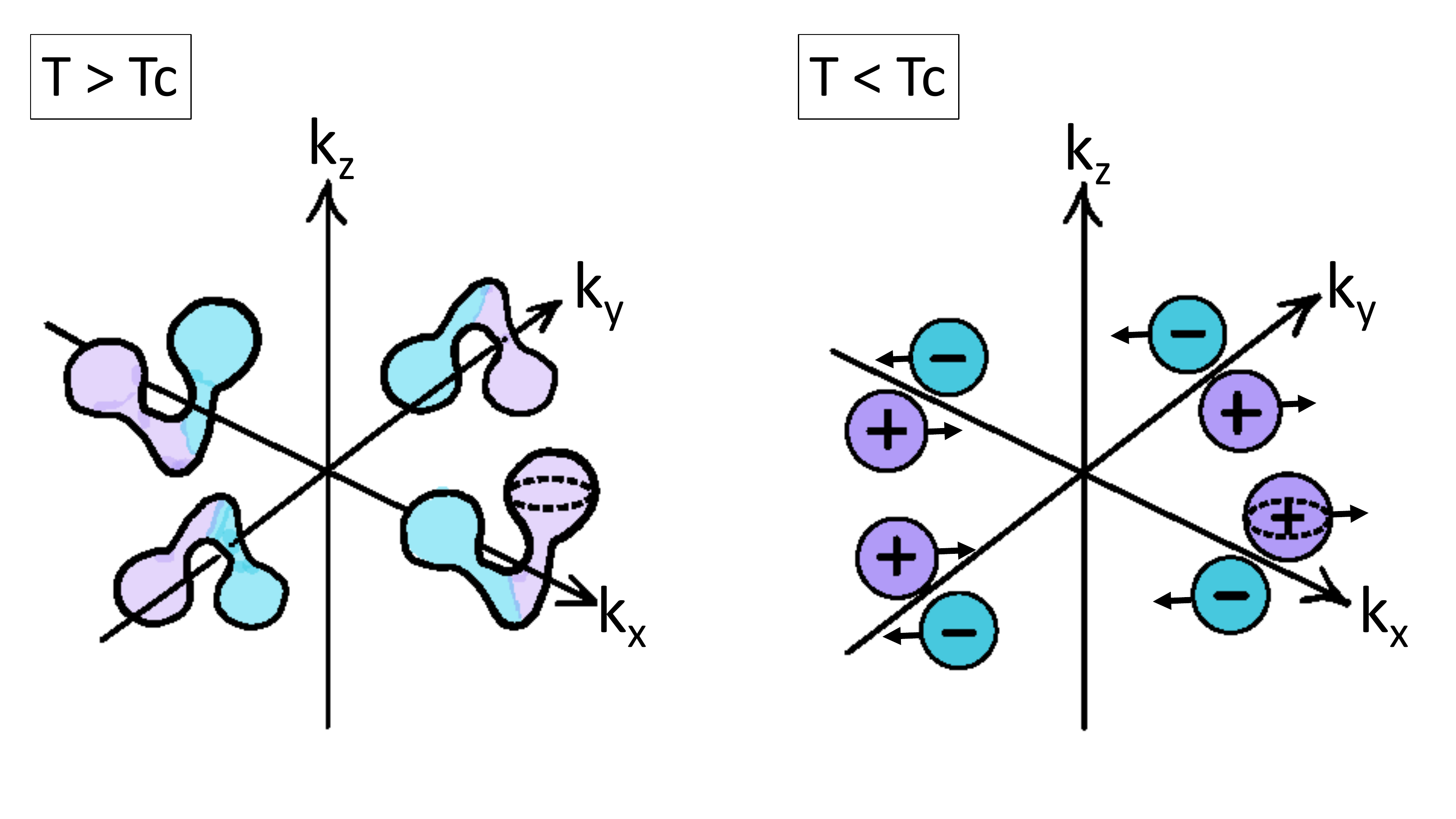}
\includegraphics[width=0.98\linewidth]{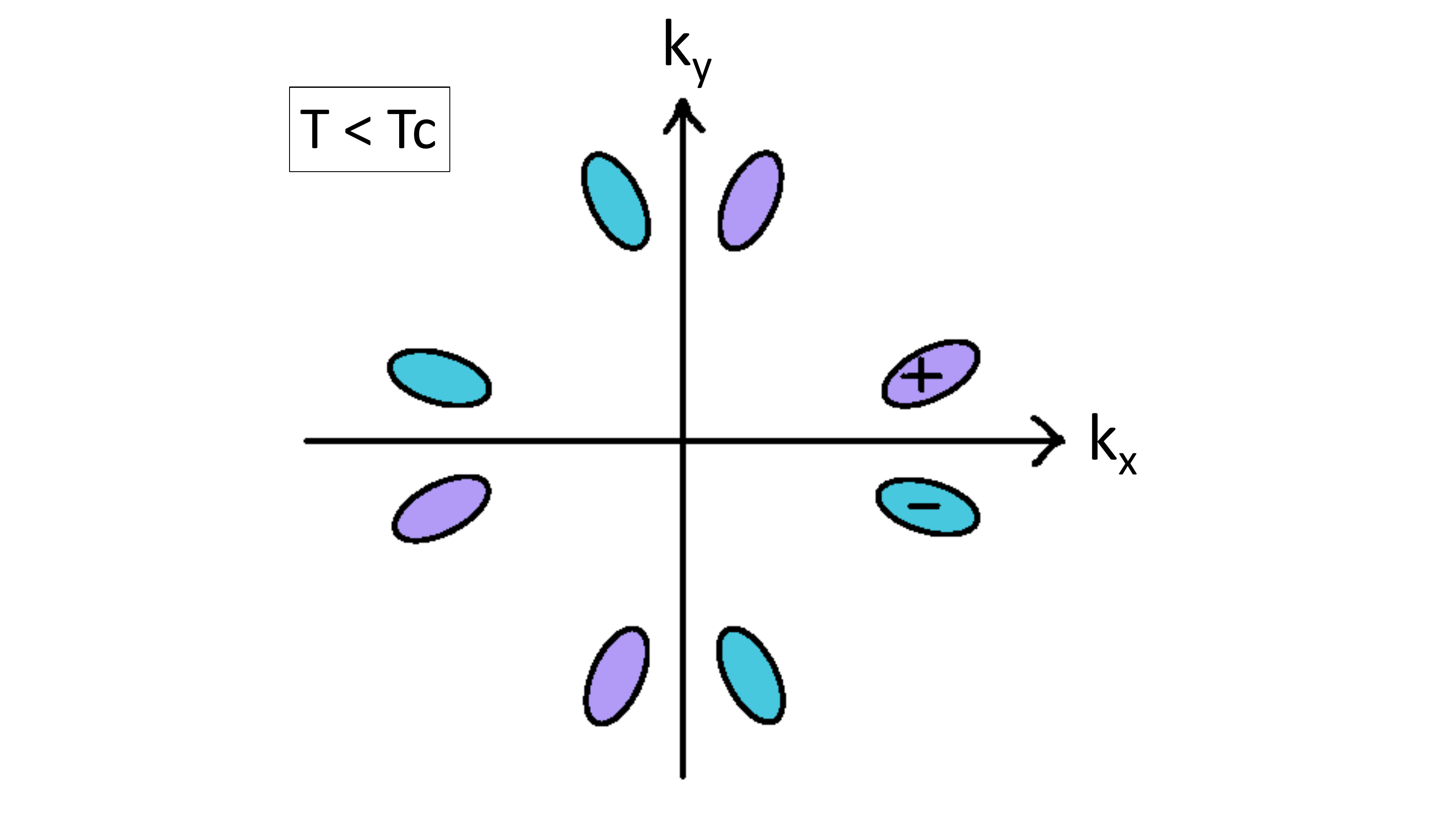}
\caption{Top: Illustrative example of $4$ topologically trivial 
Fermi surface pockets for $T \! > \! T_c$, each enclosing a pair
of Weyl points (WP) with opposite topological charge, located at momenta $(K_0,\pm K_1,0)$, $(-K_0,\pm K_1,0)$, $(\pm K_1,K_0,0)$, $(\pm K_1, -K_0,0)$.
For 
$T \!< \! T_c$, the in-plane magnetization leads to a momentum space displacement 
of the Weyl points, leading to transition into $8$ topologically nontrivial Fermi
pockets. The area of certain maximal orbits (dashed lines)
can remain unchanged across this transition. Bottom: Projected
view of the elliptical cross sections of the topologically nontrivial Fermi surfaces for $T \! < \! T_c$.}
\label{fig:fs_illustration}
\end{figure}

We choose a simple model for CeAlSi with $4$ pairs of Weyl nodes as shown in Fig.~\ref{fig:fs_illustration} (top left panel).
We have chosen the Weyl nodes and Fermi
pockets for $T > T_C$ to be consistent with the $C_{4v}$ and 
mirror $M_x, M_y$ crystal symmetries of CeAlSi, as well as 
time-reversal symmetry. These nodes crudely mimic the $W_3'$ nodes found slightly above the Fermi level in previous 
{\it ab initio} electronic structure calculations. With 
the onset of magnetism, the
Weyl nodes get displaced with opposite chirality nodes being displaced in opposite directions. As shown in Fig.~\ref{fig:fs_illustration}(top right panel), this can lead to a topological transition of the Fermi surface where each pocket now encloses a single Weyl node. At the same time, for the type of Fermi surface sketched above, certain extremal orbits can remain unchanged in area (dashed lines), so that the QO frequency will be unaffected as observed. If the Weyl nodes are not widely separated even after the topological Fermi surface phase transition, proximity to a Lifshitz transition may lead to an enhancement of the electron scattering rate (in the presence of weak disorder) due to a large density of states, which can enhance the Dingle temperature and explain the strong observed deviation from the Lifshitz-Kosevich formula.

Within the symmetry broken for $T < T_C$, it is reasonable to consider the physics of isolated Weyl nodes. We model a single Weyl node as having an elliptical pocket with velocity tensor
and a simple coupling to the Weiss field from the magnetization.
\bea
\cH_+ &=& \sigma_i G^{(+)}_{ij} \tq_j + \sigma_i M_i,\\
G^{(+)}_{ij} &=& \begin{pmatrix}
|a| & &\\
&\frac{1}{|a|} &\\
& & 1
\end{pmatrix}
\begin{pmatrix}
\cos \beta &\sin\beta & 0\\
-\sin \beta & \cos \beta & 0\\
0 & 0 & 1
\end{pmatrix},
\eea
Here $\tq_i = v_F q_i$ where $\bq$ denotes the momentum measured from the Weyl node location, and $v_F$ is a velocity scale.
The real matrix, $G_{ij}$, defined this way results in an ellipsoidal Fermi surface, whose xy-plane cross section has an elliptical shape with the major and minor axis, $|a|$ and $1/|a|$ respectively for $|a| > 1$, and the major axis is rotated from the $q_y$-axis by the angle $\beta$. For $|a| < 1$, the major axis is instead rotated from the $q_x$ axis by $\beta$.

\noindent \textbf{Position of Weyl points:} When $M_i = 0$, the Weyl point resides at $\bq = 0$. When $M_i \neq 0$, the Weyl point shifts to the point satisfying the following equation
\bea
\tq^*_i = - \left[G^{(+)}\right]^{-1}_{ij} M_j
\eea

\noindent \textbf{Eigenspectrum:} The eigenvalues of $\cH_+$ are given by
\bea
E &=& \pm \sqrt{\tq_i [G^{(+)}]^T_{ij}[G^{(+)}]_{jl}\tq_l + 2 M_i [G^{(+)}]_{ij}\tq_j + M_iM_i}\nonumber \\ 
\eea

\noindent Written in the form which is useful for numerics:

\bea
0 &=& \tq_x^2 \left(|a|^2 c^2 + \frac{1}{|a|^2}s^2\right)\nonumber\\
&&  + \tq_x \left[ 2 c s\left(|a|^2 - \frac{1}{|a|^2}\right) \tq_y + 2\left(|a|c M_x - \frac{1}{|a|} s M_y\right)\right]\nonumber\\
&& + \left[ \tq_z^2 + 2 M_z \tq_z+ \left(|a|^2s^2 + \frac{1}{|a|^2} c^2\right)\tq_y^2 \right. \nonumber \\
&& + \left. 2 \left(|a|s M_x + \frac{1}{|a|}cM_y\right)\tq_y + M^2 - E^2 \right],
\eea

\noindent where $s \equiv \sin \beta$ and $c \equiv \cos \beta$. The quadratic equation allows us to determine the mover modes given the Fermi energy and the Weiss field. If the $\tq_x$ solutions are real-valued, we obtain travelling waves; the complex-valued solutions correspond to evanescent waves.

The eigenfunctions are merely the eigenfunctions of a usual 2-by-2 Hermitian matrix, generally expressed in terms of the Pauli matrices as $d_i \sigma_i$, where $d_i = G^{(+)}_{ij} \tq_j + M_i$. The wave functions are given by
\bea
\psi &=& \begin{cases}
\frac{1}{\sqrt{2d(d+d_3)}} \begin{pmatrix}
d + d_3\\
d_1 + \I d_2
\end{pmatrix}, \text{ for } E = d > 0,\\
\frac{1}{\sqrt{2d(d+d_3)}} \begin{pmatrix}
\I d_2 - d_1\\
d_3 + d
\end{pmatrix}, \text{ for } E = -d < 0.
\end{cases}
\eea
The group velocity for a mover are given by
\bea
v_i &=& \frac{\left[G^{(+)}\right]^T_{ij} \left( \left[G^{(+)}\right]_{jl}\tq_l + M_j \right) }{E}
\eea

\noindent \textbf{Negative-chirality node:} With ${\bf M}=0$, we can use $C_{4v}$, time-reversal, and mirror symmetries 
${\cal M}_x$, ${\cal M}_y$ 
to write out the Hamiltonian for 
all $8$ Weyl nodes. For instance a negative chirality node is obtained under a mirror operation, where we can relate the g-tensor part of the Hamiltonian $\cH^{(+)}$ to the g-tensor part of $\cH^{(-)}$. For the Weyl point related to the original one by a mirror $\mathcal{M}_y$, we have
\bea
\cH^{(-)} &=& \sigma_i G^{(-)}_{ij}\tq_j + \sigma_i M_i,\\
G^{(-)} &=& - \begin{pmatrix}
|a| & & \\
& \frac{1}{|a|} & \\
& & 1
\end{pmatrix} \begin{pmatrix}
\cos \beta & - \sin \beta & 0\\
\sin \beta & \cos \beta & 0 \\
0 & 0 & 1
\end{pmatrix}.
\eea
The distinctions from $G^{(+)}$ are (i) the prefactor -1 which leads to the negative determinant and (ii) the $\sin \beta$ which used to be $-\sin \beta$ in $G^{(+)}$. The latter amounts to a rotation of the Fermi surface about $q_z$-axis by $-\beta$ instead of $\beta$ in $\cH^{(+)}$. All the formulae derived in earlier in this section can be straightforwardly generalized for $\cH^{(-)}$.

\noindent \textbf{Choice of parameters:}
As an illustrative example, we choose $|a|=0.5$ and $\beta=\pi/4$. This results in elliptical cross-sections (at any given $k_z$) for the  Fermi surfaces near a Weyl point with major : minor axis ratio of $4:1$. The major axis of the ellipse is rotated by $\pi/4$, so that it points along the $45^\circ$ direction in the $(k_x,k_y)$ plane.
We also choose other parameters to be reasonable values in line with the {\it ab initio} calculations, namely
$v_F = 500$~meV\AA and chemical potential $\mu = - 30$ meV (below the Weyl node). This leads to Fermi pockets with a typical size $k_F \sim 0.06 $~\AA$^{-1}$.
We fix the Weiss field to have a magnitude $|M| = |E_F|/4$.
We note that our results do not change qualitatively if we choose somewhat different parameters - however, it is important that the elliptical Fermi pockets are not aligned along the tetragonal $x$ or $y$ axes (see Fig.~\ref{fig:fs_illustration}).

\section*{APPENDIX F: MODELING THE DOMAIN WALL}
\label{Wall}

\begin{figure}[tb!]
	\centering
	\includegraphics[width = 0.7\columnwidth]{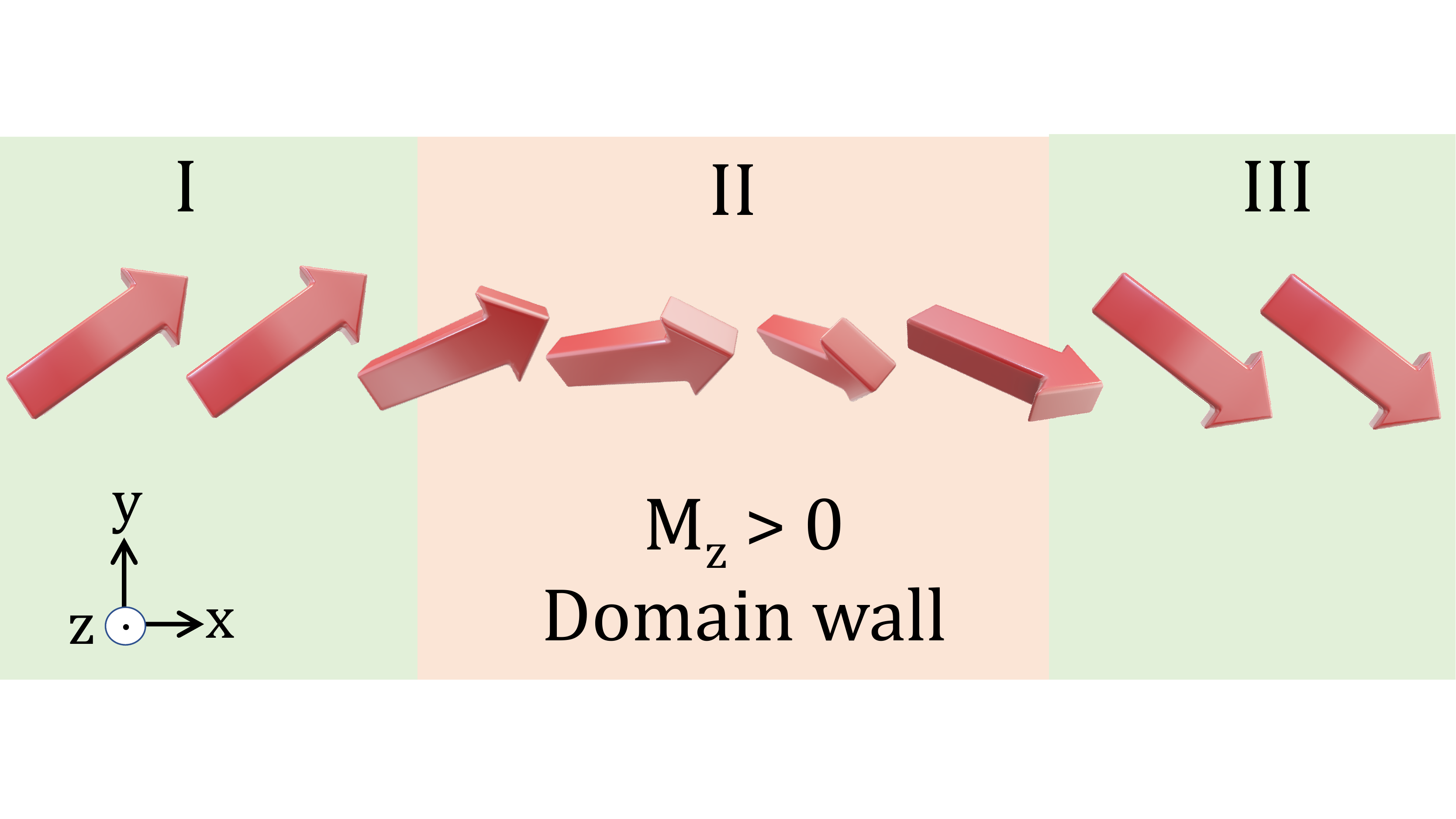}
	\caption{Evolution of the magnetization across a domain wall. Region I and region III indicate bulk domains, and region-II is the domain wall region. Going across the domain wall, the magnetization vector twists, with the perpendicular domain wall magnetization $M_z^{DW} > 0$ for the depicted configuration. We will denote ${\bf M}(x)=M (\cos\theta(x)\cos\gamma(x), \cos\theta(x)\sin\gamma(x),\sin\theta(x))$. In region-I, we choose $\theta=0,\gamma=\pi/4$, while we set $\theta=0,\gamma=-\pi/4$ in region III. In region II, we assume a twisting magnetization profile, with the maximum out of plane component determined by  $\theta_{\rm max}$ which is achieved at the center of region II.}
	\label{fig:domain}
\end{figure}

\begin{figure}[t!]
	\centering
	\includegraphics[width = 0.9\columnwidth]{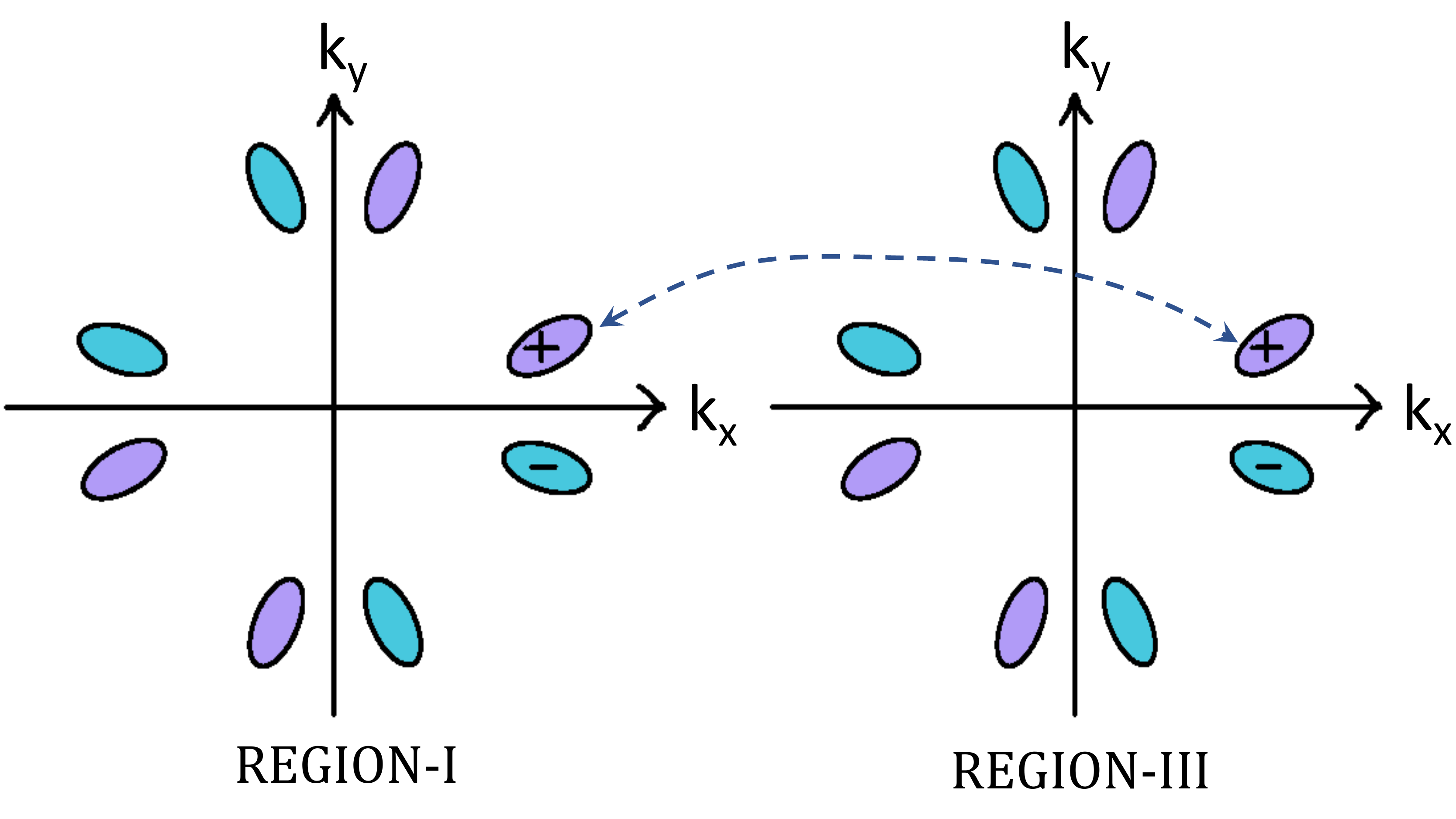}
	\caption{Illustration of the Fermi surfaces and the domain wall induced intra-node scattering (dashed arrow). For simplicity, we have not shown the displacements of the Fermi pockets relative to each other in the two domains or their difference in spin textures, but this is taken into account in our calculations as given below.}
	\label{fig:fs_scatter}
\end{figure}

We assume a domain wall width $N \times w = 40$~nm (corresponding to $N = 40$.)
We consider a domain wall between a left and a right region (see illustration in Fig.~\ref{fig:domain}) with the following Weiss fields, respectively,
\bea
\bM_I &=& M \left(\cos \gamma, \sin \gamma, 0\right)^T,\\
\bM_{III} &=& M \left(\cos \gamma, -\sin \gamma, 0\right)^T.
\eea
The domain wall region, region II, has a width $N w$, which is partitioned into $N$ intervals, each with width $w$. The Weiss field in $j$-th interval is given by
\bea
\bM_j &=& M \left(\cos \theta_j \cos \gamma_j, \cos \theta_j \sin \gamma_j, \sin \theta_j\right)^T, \\
\gamma_j &=& \gamma - 2 \gamma\frac{j - 1/2}{N},\\
\theta_j &=& \left(1 - \frac{2|j - \frac{1}{2} - \frac{N}{2}|}{N}\right) \theta,
\eea
where $\theta$ is the angle at the center of region II. $\theta_j$ monotonically decreases away from the center of region II. A large $N$ models a smooth variation of the Weiss field in region II.

\subsection{Transmission and reflection coefficients}

The domain wall will lead to a scattering between Weyl Fermi surfaces. For simplicity, we assume a smooth domain wall and only take into account the intra-node scattering as shown in Fig.~\ref{fig:fs_scatter}.
We now sketch the computation of the transmission coefficient (TC) and reflection coefficient (RC) at the domain wall, defined earlier. For concreteness, we show the calculation for $\cH^{(+)}$. The step-by-step summary is given below
\begin{itemize}
	\item For a given Fermi energy $E$ and the parallel momenta $(q_y, q_z)$, we compute the x-momenta for all the regions.
	\item Compute the eigenfunctions
	\item Wave function in each region is a superposition of a left mover and a right mover, except in Region III, where the wave function consists of only a right mover
	\bea
	\Psi_I &=& \chi_R e^{\I \bq_R \cdot \bx} + r\chi_Le^{\I \bq_L \cdot \bx} \\
	\Psi_{II, j} &=& c^{(j)}_1 \eta_1^{(j)} e^{\I \bp^{(j)}_1 \cdot \bx} + c^{(j)}_2 \eta_2^{(j)} e^{\I \bp^{(j)}_2 \cdot \bx},\\
	\Psi_{III} &=& t \xi_R e^{\I \bk_R\cdot \bx},
	\eea
	where $t$ and $r$ are the transmission and reflection amplitude respectively.
	\item We then match the wave function at each boundary at $x_j = j w$ for $j = 0, 1, \cdots, N$ to determine $r, t$ and $c^{(j)}_{1, 2}$. This can be formulated in transfer matrix form. This can be seen below.
	\end{itemize}
	\begin{widetext}
	\bea
	\chi_R + r \chi_L &=& c_1^{(1)} \eta_1^{(1)} + c_2^{(1)} \eta_2^{(1)},\\
	c_1^{(1)} \eta_1^{(1)} e^{\I p_{1x}^{(1)} w} + c_2^{(1)} \eta_2^{(1)} e^{\I p_{2x}^{(1)} w} &=& c_1^{(2)} \eta_1^{(2)} e^{\I p_{1x}^{(2)} w} + c_2^{(2)} \eta_2^{(2)} e^{\I p_{2x}^{(2)} w}\\
	c_1^{(2)} \eta_1^{(2)} e^{\I p_{1x}^{(2)} 2w} + c_2^{(2)} \eta_2^{(2)} e^{\I p_{2x}^{(2)} 2w} &=& c_1^{(3)} \eta_1^{(3)} e^{\I p_{1x}^{(3)} 2w} + c_2^{(3)} \eta_2^{(3)} e^{\I p_{2x}^{(3)} 2w}\\
	c_1^{(j)} \eta_1^{(j)} e^{\I p_{1x}^{(j)} jw} + c_2^{(j)} \eta_2^{(j)} e^{\I p_{2x}^{(j)} jw} &=& c_1^{(j+1)} \eta_1^{(j+1)} e^{\I p_{1x}^{(j+1)} jw} + c_2^{(j+1)} \eta_2^{(j+1)} e^{\I p_{2x}^{(j+1)} jw}\\
	c_1^{(N)} \eta_1^{(N)} e^{\I p_{1x}^{(N)} Nw} + c_2^{(N)} \eta_2^{(N)} e^{\I p_{2x}^{(N)} Nw} &=& t \xi_R e^{\I k_{Rx} Nw}.
		\eea
	\noindent Rewriting in matrix form
	\bea
	\begin{pmatrix}
	\chi_R & \chi_L
	\end{pmatrix} \begin{pmatrix}
	1 \\
	r
	\end{pmatrix} &=& \begin{pmatrix}
	\eta_1^{(1)} & \eta_2^{(1)}
	\end{pmatrix}
	\begin{pmatrix}
	c_1^{(1)}\\
	c_2^{(1)}
	\end{pmatrix},\\
	\begin{pmatrix}
	\eta_1^{(j)} e^{\I p_{1x}^{(j)} j w} & \eta_2^{(j)} e^{\I p_{2x}^{(j)} j w} 
	\end{pmatrix} \begin{pmatrix}
	c^{(j)}_1 \\
	c^{(j)}_2
\end{pmatrix}	 &=& \begin{pmatrix}
	\eta_1^{(j+1)} e^{\I p_{1x}^{(j+1)} j w} & \eta_2^{(j+1)} e^{\I p_{2x}^{(j+1)} j w} 
	\end{pmatrix} \begin{pmatrix}
	c^{(j+1)}_1 \\
	c^{(j+1)}_2
\end{pmatrix}\\
	\begin{pmatrix}
	\eta_1^{(N)} e^{\I p_{1x}^{(N)} N w} & \eta_2^{(N)} e^{\I p_{2x}^{(N)} N w} 
	\end{pmatrix} \begin{pmatrix}
	c^{(N)}_1 \\
	c^{(N)}_2
\end{pmatrix} &=& t \xi_R e^{\I k_{Rx} Nw}.
	\eea

\noindent From above, we can solve for $r$ and $t$ by the transfer matrix $T^{(j)}$:
\bea
\begin{pmatrix}
1\\
r
\end{pmatrix} &=& t e^{\I k_{Rx}Nw} \begin{pmatrix}
\chi_R & \chi_L
\end{pmatrix}^{-1} T^{(1)} \cdots T^{(N)} \xi_R \equiv t \begin{pmatrix}
u_1 \\
u_2 
\end{pmatrix}\nonumber \\ \\
T^{(j)} &=& \begin{pmatrix}
\eta_1^{(j)} e^{\I p_{1x}^{(j)}(j-1)w} & \eta_2^{(j)} e^{\I p_{2x}^{(j)}(j-1)w}
\end{pmatrix}
\begin{pmatrix}
\eta_1^{(j)} e^{\I p_{1x}^{(j)}jw} & \eta_2^{(j)} e^{\I p_{2x}^{(j)}jw}
\end{pmatrix}^{-1},
\eea
\end{widetext}

\noindent where, in the definition of the transfer matrix, the two matrices differ from each other, apart from the inverse operation, by the phase factors: one involves $(j-1)w$, whereas the other involves $jw$. We finally obtain
\bea
t &=& 1/u_1\\
r &=& u_2/u_1.
\eea
TC and RC are given by
\bea
TC &=& \frac{|v_{x, \text{trans}}|}{|v_{x, \text{inc}}|} |t|^2,\\
RC &=& \frac{|v_{x, \text{refl}}|}{|v_{x, \text{inc}}|} |r|^2.
\eea

The longitudinal conductance $g_{xx}$ and the transverse conductance $g_{yx}$ are then computed using TC and RC \cite{sorn2021domain} (see also Refs. \onlinecite{REF2} and \onlinecite{REF3}.)

\subsection{Results: case $|a| = 0.5, \beta = \pi/4$}

\subsubsection{Parameters}
The $xy$-plane cross section of the Fermi surface near a Weyl point is an ellipse whose major axis is rotated by $\pi/4$.

The results corresponds to the following parameters:
\begin{itemize}
	\item Fermi velocity $v_F = 500$~meV.\AA
	\item Fermi wave vector $k_F \sim 0.06$~\AA$^{-1}$
	\item Fermi energy $E_F = - 30$~meV
	\item Weiss field $|M| = 0.015$~a.u., which corresponds to $|E_F|/4$.
	\item Domain wall width $N \times w = 40$~nm  for $N = 40$. 
\end{itemize}

\subsubsection{Results}

\begin{figure}[!b]
	\centering
	\includegraphics[width = 0.9\columnwidth]{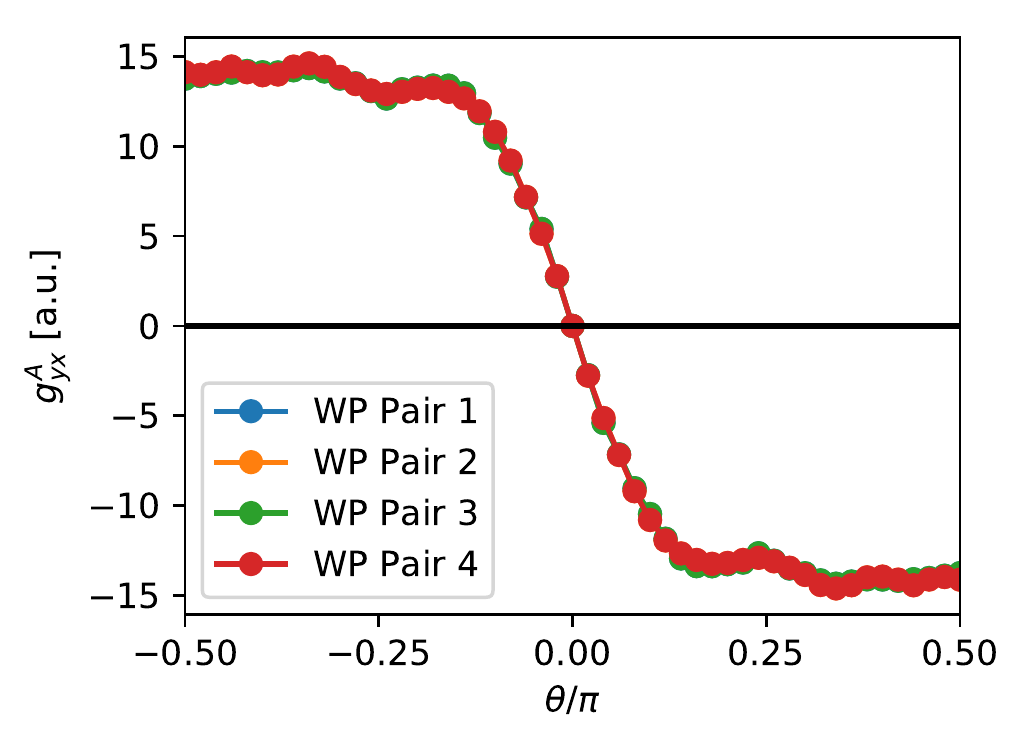}
	\caption{Contributions from the 4 pairs (green-purple pair of adjacent Fermi surfaces related by $\mathcal{M}_x$ or $\mathcal{M}_y$ in Fig.~S\ref{fig:fs_illustration}) of Fermi surfaces.}
	\label{fig:4wps}
\end{figure}

\begin{figure}[!tbh]
	\centering
	\includegraphics[width=0.9\columnwidth]{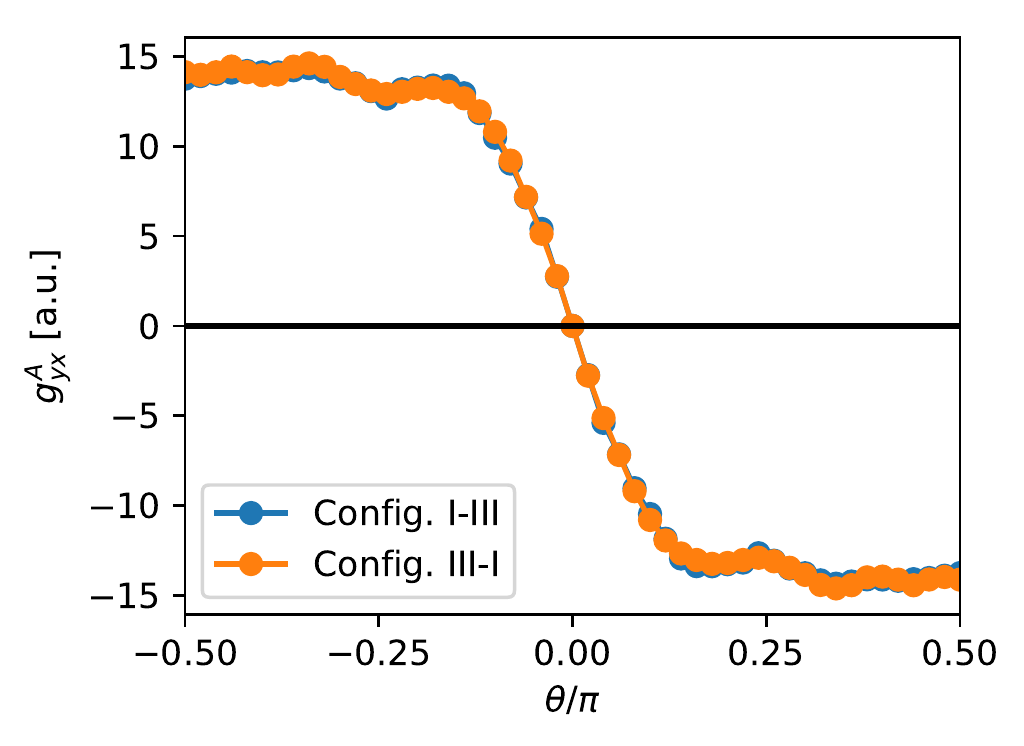}
	\caption{$\theta$ dependence of AHE stays the same upon interchanging Region I and Region III.}
	\label{fig:LR_vs_RL}
\end{figure}


We will show results of the anomalous Hall contribution obtained by antisymmetrizing the off-diagonal conductance: $g^A_{yx}(\theta)=\frac{1}{2}(g_{yx}(\theta) - g_{yx}(-\theta))$, namely antisymmetrizing w.r.t merely reversing the $M_z$ component of the Weiss field. Figure~\ref{fig:4wps} shows $g^A_{yx}$ for the 4 pairs of WPs: (i) Fermi surfaces in each pair are related by either $\mathcal{M}_x$ or $\mathcal{M}_y$ mirror operation at zero Weiss field (see Fig.~\ref{fig:fs_illustration}), and (ii) different pairs are related by a $C_{4v}$ rotation at zero Weiss field (see Fig.~\ref{fig:fs_illustration}.)

A few main results are summarized below:
\begin{itemize}
	\item[(1)] AHE contributions from the two Fermi surfaces in each pair has opposite signs, yet they do not cancel each other out, so AHE is still non vanishing.
	\item[(2)] AHE from the four WPs related by $C_{4z}$ rotations has the same sign (see Fig.~\ref{fig:4wps}.)
	\item[(3)] Interchanging region I and region III leads to the same $\theta$-dependence of $g^A_{yx}$ (see Fig.~ \ref{fig:LR_vs_RL}.) This suggests that as long as the total $z$-component of the Weiss field does not vanish, AHE contribution from the domain wall is non-zero.
	\item[(4)] Ratio $g^A_{yx}/g_{xx}$ is of the order $10^{-3}$ at small angle, see Fig.~\ref{LHE_WeylScatt}(c).
\end{itemize}

\bibliography{CeAlSi}

\end{document}